%% file: melba-sample.tex
\documentclass[twoside,11pt]{article}

\usepackage[accepted]{melba}

\usepackage{amsmath,amsfonts}
\usepackage{bm}
\usepackage{multirow}
\usepackage{multicol}
\usepackage{layouts}
\usepackage{float}
\usepackage{graphicx, array,booktabs}
\usepackage{soul}
\usepackage{amsmath,amsfonts}
\usepackage{letltxmacro}

\LetLtxMacro{\originaleqref}{\eqref}
\renewcommand{\eqref}{Eq.~\originaleqref}
\newcolumntype{N}{>{\centering\arraybackslash}m{.5in}}
\newcolumntype{G}{>{\centering\arraybackslash}m{\dimexpr2in+6\tabcolsep}}
\newcolumntype{P}[1]{>{\centering\arraybackslash}p{#1}}

\newcommand{\best}[1]{\textcolor{blue}{#1}}
\newcommand{\secbest}[1]{\textcolor{red}{#1}}

\melbaid{2024:011}  
\doi{https://doi.org/10.59275/j.melba.2024-88fa}
\melbaauthors{Qian Tao}  
\volume{2}
\firstpageno{856}  
\melbayear{2024}  
\datesubmitted{02/2024}  
\datepublished{06/2024}  

\melbaspecialissue{TO BE DEFINED}
\melbaspecialissueeditors{TO BE DEFINED}

\ShortHeadings{Bayesian uncertainty estimation by HMC in cardiac MRI}{Y. Zhao, J. Tourais, I. Pierce, T.A. Treibel, S. Weing\"artner, A.M. Schweidtmann, and Q. Tao}

\title{Bayesian uncertainty estimation by Hamiltonian Monte Carlo: Applications to cardiac MRI segmentation}

\author{\firstname Yidong \surname Zhao \orcid{0000-0003-3953-6921} \email Y.Zhao-8@tudelft.nl \\  
	\addr Department of Imaging Physics, Delft University of Technology, Delft, The Netherlands
        \AND 
        \firstname Jo{\~a}o \surname Tourais \orcid{0000-0002-1388-4023} \email J.L.SilvaCanaveiraTourais@tudelft.nl \\
	\addr Department of Imaging Physics, Delft University of Technology, Delft, The Netherlands
        \AND 
        \firstname Iain \surname Pierce \email iain.pierce@nhs.net \\
	\addr Barts Heart Centre, Barts Health NHS Trust, London, United Kingdom \\ Institute of Cardiovascular Science, University College, London, United Kingdom
        \AND 
        \firstname Christian \surname Nitsche \email christian.nitsche@nhs.net \\
	\addr Barts Heart Centre, Barts Health NHS Trust, London, United Kingdom \\ Institute of Cardiovascular Science, University College, London, United Kingdom
        \AND 
        \firstname Thomas A. \surname Treibel \email thomas.treibel@nhs.net \\
	\addr Barts Heart Centre, Barts Health NHS Trust, London, United Kingdom \\ Institute of Cardiovascular Science, University College, London, United Kingdom
        \AND 
        \firstname Sebastian \surname Weing{\"a}rtner \orcid{0000-0002-0739-6306} \email S.Weingartner@tudelft.nl \\
	\addr Department of Imaging Physics, Delft University of Technology, Delft, The Netherlands
        \AND 
        \firstname Artur M. \surname Schweidtmann \orcid{0000-0001-8885-6847} \email A.Schweidtmann@tudelft.nl \\
	\addr Department of Chemical Engineering, Delft University of Technology, Delft, The Netherlands
        \AND 
        \firstname Qian \surname Tao \orcid{0000-0001-7480-0703} \email Q.Tao@tudelft.nl \\
	\addr Department of Imaging Physics, Delft University of Technology, Delft, The Netherlands
}

\begin{document}

\maketitle

\begin{abstract}
Deep learning (DL)-based methods have achieved state-of-the-art performance for many medical image segmentation tasks. Nevertheless, recent studies show that {deep neural networks (DNNs)} can be miscalibrated and overconfident, {leading to "silent failures'' that are risky} for clinical applications. Bayesian DL provides an intuitive approach to DL failure detection, based on posterior probability estimation.  However, the posterior is intractable for large medical image segmentation DNNs. To tackle this challenge, we propose a Bayesian learning framework using Hamiltonian Monte Carlo (HMC), tempered by cold posterior (CP) to accommodate medical data augmentation, named HMC-CP. For HMC computation, we further propose a cyclical annealing strategy, capturing both local and global geometries of the posterior distribution, enabling highly efficient Bayesian DNN training with the same computational budget as training a single DNN. The resulting Bayesian DNN outputs an ensemble segmentation along with the segmentation uncertainty. We evaluate the proposed HMC-CP extensively on cardiac magnetic resonance image (MRI) segmentation, using in-domain steady-state free precession (SSFP) cine images as well as out-of-domain datasets of quantitative $T_1$ and $T_2$ mapping. Our results show that the proposed method improves both segmentation accuracy and uncertainty estimation for in- and out-of-domain data, compared with well-established baseline methods such as Monte Carlo Dropout and Deep Ensembles. Additionally, we establish a conceptual link between HMC and the commonly known stochastic gradient descent (SGD) and provide general insight into the uncertainty of DL. This uncertainty is implicitly encoded in the training dynamics but often overlooked. With reliable uncertainty estimation, our method provides a promising direction toward trustworthy DL in clinical applications. \footnote{{We release our code in} \url{https://gitlab.tudelft.nl/yidongzhao/hmc_uncertainty}.}
\end{abstract}
%
%
%
\section{Introduction}
\label{sec:introduction}
Image segmentation is an integral part of medical image post-processing in a wide range of clinical applications~\citep{chen2020deep}. However, the manual delineation of anatomical features or organs is a demanding and time-consuming task in clinical practice. Deep learning (DL)-based automatic segmentation methods, in particular the U-Net and its variants~\citep{ronneberger2015u,isensee2021nnu}, have demonstrated excellent performance in automated medical image segmentation and become the \textit{de facto} standard~\citep{bernard2018deep,campello2021multi} in literature. Nonetheless, the robustness and reliability of {deep neural networks (DNN)} remain a major concern for clinical use when tested on data with domain shift~\citep{campello2021multi,yan2019domain,2020MRI}. Ideally, such uncertainty can be indicated by the {Softmax score~{\citep{guo2017calibration}}}. However, recent studies found DNNs to be seriously miscalibrated~\citep{guo2017calibration,minderer2021revisiting,wang2021rethinking}, i.e., the confidence score provided by the Softmax output does not match the empirical accuracy~\citep{mehrtash2020confidence}. While tested on unseen, heterogeneous data, the DL models often output high confidence Softmax score in erroneous predictions, leading to ``silent failures''~\citep{gonzalez2021detecting}. This severely undermines the trustworthiness of DL models to clinicians and patients and causes high risks for clinical applications. Therefore, accurate uncertainty estimation, i.e., reporting low confidence when an error likely occurs, has important clinical implications on the trustworthiness of DL systems in real-world deployment~\citep{jungo2019assessing}. 

\subsection{Related Work}
Uncertainty in medical imaging segmentation has recently moved into the attention of the community~\citep{kohl2018probabilistic, baumgartner2019phiseg,wang2019aleatoric, jungo2019assessing, jungo2020analyzing, gonzalez2021detecting, mehrtash2020confidence}. Previous work has investigated two different types of uncertainty. Part of the work focused on the intrinsic ambiguity of contour definition, inherent to the difficulty of segmentation tasks, which is referred to as the \emph{aleatoric} uncertainty~\citep{hora1996aleatory,der2009aleatory}. This uncertainty cannot be reduced by collecting more data~{\citep{hullermeier2021aleatoric}}. Kohl \emph{et~al.} proposed a Probabilistic U-Net~\citep{kohl2018probabilistic} which models the variation among experts in manual contouring and aims at generating various feasible segmentation masks to estimate the uncertainty. Baumgartner~\emph{et~al.} proposed PHi-Seg, which assumes that the segmentation map is {intrinsically ambiguous} and governed by hierarchical latent features, while probabilistic predictions can be made via sampling from the learned latent feature distribution~\citep{baumgartner2019phiseg}. Test time augmentation~(TTA){~\citep{wang2019aleatoric}} was also proposed to estimate the aleatoric uncertainty of contours via averaging predictions on augmented inputs. {A more advanced technology proposed by {\cite{ouyang2022improved}} combines TTA, adaptive temperature scaling, and shape feasibility.} However, the aleatoric uncertainty reflects intrinsic ambiguity of the task rather than indicating failures of trained networks.

Another type of uncertainty is the \emph{epistemic} uncertainty that reflects the model uncertainty when tested on heterogeneous data.~\citep{hullermeier2021aleatoric, hora1996aleatory, der2009aleatory, kendall2017uncertainties}. Most of the epistemic estimation methods fall into the Bayesian neural network (BNN) framework~\citep{mackay1995bayesian,lampinen2001bayesian,wang2020survey,marcot2019advances}. {The BNNs model the posterior probability of the learned DNN weights~{\citep{mackay1995bayesian, neal2012bayesian}}}. The predictive uncertainty is inferred from the distribution of the DNN weights and subsequently that of the DNN predictions{~\citep{gal2016dropout, kendall2017uncertainties, mehrtash2020confidence}}. {Modern BNN architectures also learn the distribution of explainable variables that govern the mapping from images to segmentations instead of weights for better generalization~{\citep{gao2023bayeseg}}}. However, the posterior distribution is prohibitively difficult to analytically derive for large networks{~\citep{blundell2015weight, blei2017variational, neal2012bayesian}}. As such, the Variational Inference~(VI) method~\citep{blei2017variational} has been proposed for the posterior approximation, which models network weights as independent Gaussian random variables~\citep{blundell2015weight}. Such approximation is intrinsically limited by the strong assumption of Gaussian posterior. Moreover, it practically doubles {the network's parameters} and can become unstable during training~\citep{ovadia2019can, jospin2022hands}. The Monte-Carlo Dropout~(MC-Dropout)~\citep{gal2016dropout,kendall2017uncertainties} proposed by Gal \textit{et al.} can be considered as a VI proxy, assuming that the posterior of weights is modulated by a random Bernoulli random variable. In the same spirit, Bayesian SegNet~\citep{kendall2015bayesian} employed dropout layers in the bottleneck layers of fully convolutional networks for uncertainty estimation. 

Unfortunately, uncertainty estimation by VI and its MC-Dropout proxy remained insufficient for large DL models. Recent work reported silent failures, poor calibration, and degraded segmentation performance~\citep{folgoc2021mc,gonzalez2021self,gonzalez2021detecting}. Fort \emph{et al.} showed that the VI-based methods including MC-Dropout only explore a limited, local weight space due to the restrictive assumptions. In comparison, Deep Ensembles~\citep{lakshminarayanan2016simple, mehrtash2020confidence} estimate network uncertainty via averaging independently trained network instances. The independently trained instances can be seen as a combination of maximum-a-posteriori~(MAP) solutions~\citep{fort2019deep}. The ability to globally explore solutions makes Deep Ensembles the best-performing uncertainty estimation method so far~{\citep{ovadia2019can,abdar2021review,fort2019deep,gustafsson2020evaluating}}. However, theoretical and practical limitations remain: first, Deep Ensembles ignores the local posterior geometry around the MAP solution, which was reported to be important for DNN calibration~\citep{garipov2018loss, maddox2019simple, mingard2021sgd}; second, the time complexity of Deep Ensembles grows linearly with the number of models. It becomes computationally prohibitive, given that training a single network is time-and-energy-consuming. 

In this work, we aim to address the aforementioned limitations of BNNs for medical image segmentation: the VI and MC-Dropout methods have limited approximation capacity, while Deep Ensembles fail to cover local posterior distribution while being computationally inefficient. We propose to use the Markov Chain Monte Carlo~(MCMC)~\citep{hammersley2013monte, hastings1970monte} approach, and in particular the Hamiltonian Monte Carlo~(HMC)~\citep{neal2011mcmc,chen2014stochastic}. HMC treats sampling of a target distribution as modeling of particle motion~\citep{risten1989fokker,sarkka2019applied}. It is theoretically guaranteed that simulating the Hamiltonian dynamics yields samples conforming to the target distribution~\citep{neal2011mcmc}, hence it theoretically promises improved BNN uncertainty estimation compared with previous methods with restrictive assumptions. 

{{In literature, \cite{izmailov2021bayesian}} employed a full-batch HMC to explore the precise posterior of neural networks. However, the full-batch HMC is not scalable to large neural networks because of the computational efficiency~{\citep{izmailov2021bayesian}}. The first scalable HMC with stochastic gradient (SGHMC) on neural networks was proposed by~{\cite{chen2014stochastic}} for posterior estimation. Further works reveal that tempering the posterior is needed for HMC sampling with stochastic gradient~{\citep{zhang2019cyclical,wenzel2020good}}}. {However, these early attempts~{\citep{chen2014stochastic,izmailov2018averaging,zhang2019cyclical}} focus on simple classification or regression tasks. The research on segmentation networks' behavior with a dense output under the HMC dynamics is seriously limited, and warrants further investigation. For example, {\cite{wenzel2020good}} reported the necessity of a tempered posterior in the classification tasks but {\cite{izmailov2021bayesian}} claimed that it is not necessary for the full-batch HMC but rather an artifact of data augmentation. This raises questions on the posterior choice in Bayesian segmentation networks, where data augmentation is heavily used.} A standard method of posterior distribution modeling is to use Gibbs distribution~\citep{lifshitz1984statistical} and treat the inverse of the predefined loss function as ``energy"~\citep{carvalho2020scalable,kapoor2022uncertainty}. However, training data augmentation, as is commonly used in medical image applications due to data scarcity, would render the exact modeling of posterior intractable, as the independent number of observed data samples becomes ambiguous after data augmentation. This leads to the so-called ``dirty likelihood'' effect and results in degraded performance of BNNs~\citep{nabarro2021data, wenzel2020good}. {Moreover, the sampling strategy of the HMC chain remains understudied in segmentation networks, including e.g. the number of HMC samples needed for proper calibration.} IN this work, we propose to investigate and evaluate cold posterior in Bayesian segmentation to research the ``dirty likelihood" in the presence of data augmentation. 

Additionally, a largely unanswered question is whether the diversity in the posterior weight space $W$ propagates to that of the functional space $f_W(\cdot)$. We differentiate these two spaces because there is no simple relationship between the two due to symmetry\footnote{A permuted set of weights, for example, can lead to the same function.}, whereas functional space diversity critically determines the quality of uncertainty estimation~\citep{kendall2017uncertainties, fort2019deep}. Limited research has been done to investigate the functional diversity of BNNs, and none for medical image segmentation applications. For classification, Fort \emph{et al.} used cosine similarity to analyze the similarity between posterior weights and evaluated the predictive diversity in functional space via comparing classification agreement~\citep{fort2019deep}. Segmentation networks, however, have much more complicated output in high dimensions~\citep {ronneberger2015u}. In this work, we propose to evaluate the functional space diversity of BNNs for segmentation uncertainty beyond that of the posterior weight space. 

{In practice, the out-of-domain performance of uncertainty estimation is crucial to failure detection in cardiac MRI because of the domain shifts caused by imaging protocol variations. Previous works~{\citep{ovadia2019can,izmailov2021bayesian}} researched the network behavior under simulated distortions such as additive noise and blurring. However, real-world domain shift appears much more complicated than such in-silico distortion simulations, such as novel quantitative MRI mapping versus conventional cine MRI.} The performance of image segmentation and uncertainty estimation under such drastic domain shifts have not been systematically studied. Finally, for ease of use in clinical practice and scalable data analysis~\citep{jungo2019assessing,czolbe2021segmentation}, an aggregated confidence score that can detect the segmentation failure on the image level would be highly clinically relevant, which obviates the need for users to review the voxel-level uncertainty maps~\citep{kendall2015bayesian,czolbe2021segmentation}. 

\subsection{Contributions}
This study substantially extends the theory, analysis, and application of our previous work published in MICCAI 2022~\citep{zhao2022efficient}, in which we proposed the training checkpoint ensemble during SGD with momentum. In particular, we developed the theoretical foundation of HMC uncertainty estimation and absorbed the previously proposed method as a special case. Specifically, we have made the following contributions:
\begin{itemize}
    \item We propose a Bayesian DL framework for medical image segmentation using HMC-CP, which delivers improved uncertainty estimation compared with state-of-the-art baseline methods. The proposed method is highly efficient in computation with the novel annealing learning strategy for multi-modal posterior sampling, {because of the natural resemblance between HMC sampling and SGD optimization. We systematically investigated the effect of cold-posterior in the cardiac MRI segmentation network and researched the calibration performance with various numbers of posterior samples}. 
    
    \item We extensively analyze the functional diversity of the Bayesian segmentation networks by the proposed and other existing methods. We demonstrate that the proposed method yields superior functional diversity compared with other methods, leading to more accurate uncertainty estimation.
    \item We propose an image-level uncertainty score for ease of use in clinical practice and evaluated our proposed method on datasets covering substantial domain shifts from cine to quantitative MRI data. Empirical results showed that the proposed score can effectively detect segmentation failure, for both in-domain and out-of-domain datasets.
\end{itemize}

\section{Methods}
\subsection{Posterior Modelling of Segmentation Networks}
\label{sec:post_modelling}
{BNNs admit a statistical model} $p(\bm{w})$ as its prior distribution over the network weights~\citep{jospin2022hands}, which characterizes the weight distribution before observing any data. Following~\citep{carvalho2020scalable,hammam2021real}, we assume the weight prior as a zero-mean Gaussian: $p\left(\bm{w}\right)\sim \mathcal{N}\left(0, \frac{1}{\lambda} \mathbb{I} \right)$, where $\lambda$ controls the prior variance. According to Bayes' Theorem, the weight distribution can be re-estimated after observing the dataset, known as posterior~\citep{neal2012bayesian}. In this section, we define the posterior distribution of segmentation models. 

\subsubsection{The Weight Posterior} Given a training dataset $\mathcal{D}=\left\{\bm{x}_i, \bm{y}_i\right\}_{i=1}^{n}$ with $n$ image-label pairs, the training procedure learns a weight setting $\bm{w}$ that minimizes the discrepancy between $f_{\bm{w}}(\bm{x}_i)$ and $\bm{y}_i$ for $i=1, 2, \dots, n$, where $f_{\bm{w}}$ is the DNN parameterized by $\bm{w}$. The prior distribution of weights shrinks to the posterior with the presence of $\mathcal{D}$. According to Bayes' Theorem, the following relationship holds:

\begin{equation}
    p\left(\bm{w}|\mathcal{D}\right) \propto p\left(\mathcal{D}|\bm{w}\right)\cdot p\left(\bm{w}\right)
\end{equation} 
where the likelihood term $p\left(\mathcal{D}|\bm{w}\right)$ measures how well the network prediction $f_{\bm{w}}(\bm{x})$ on a training sample $\bm{x}$ with weight $\bm{w}$ aligns with the ground truth $\bm{y}$ in the training set.

In this work, we adopt the widely-used nnU-Net~\citep{isensee2021nnu} as our $f_{\bm{w}}$, and use a combination of soft-Dice loss $\mathcal{L}_{DSC}$ and cross entropy loss $\mathcal{L}_{\mathrm{CE}}$ to estimate the discrepancy between the network prediction $f_{\bm{w}}(\bm{x})$ and the ground truth $\bm{y}$. For an image with $N$ voxels, let $p(\hat{\bm{y}}_i=c|\bm{x}_i, {w})$ be the predictive probability of voxel $\bm{x}_i$ belonging to class $c$ with $C$ semantic classes in total, the soft-Dice loss is defined as:
\begin{equation}
    \mathcal{L}_\mathrm{DSC} = -2 \sum_{c=1}^{C}\frac{\sum_{i=1}^{N}p(\hat{\bm{y}}_i=c|\bm{x}_i, \bm{w}) \cdot (\bm{y}_i = c)}{\sum_{i=1}^{N}p(\hat{\bm{y}}_i=c|\bm{x}_i, \bm{w}) + (\bm{y}_i = c)},
\end{equation}
and the cross entropy loss is defined as:
\begin{equation}
    \mathcal{L}_{\mathrm{CE}} = -\frac{1}{N} \sum_{i=1}^{N}\sum_{c=1}^{C} \log p(\hat{\bm{y}}_i=c|\bm{x}_i, \bm{w}) \cdot (\bm{y}_i = c).
\end{equation}
The total loss $\mathcal{L}(\bm{w}) = \mathcal{L}_\mathrm{DSC}(\bm{w}) +  \mathcal{L}_\mathrm{CE}(\bm{w})$ measures how likely the training samples are observed under the weight setting $\bm{w}$. In this work, we follow~\citep{carvalho2020scalable,wenzel2020good} and define the likelihood with Gibbs distribution~\citep{lifshitz1984statistical}:
\begin{equation}
\begin{split}
    \label{eq:likihood}
    p\left(\mathcal{D}|\bm{w}\right) \propto \exp{\left[ -\sum_{i=1}^{n}  \mathcal{L} \left(f_{\bm{w}}(\bm{x}_i), \bm{y}_i \right)\right]}.
\end{split}
\end{equation}
{We aim to draw samples that maximize the log-posterior $\log p\left(\bm{w}|\mathcal{D}\right)$, which is equivalent to minimizing the following energy function during training}:
\begin{equation}
\begin{split}
\label{eq:loss_function}
    U(\bm{w}) &= -\log p\left(\bm{w}|\mathcal{D}\right) = - \log p\left(\mathcal{D}|\bm{w}\right)  - \log p \left(\bm{w}\right) \\
    &\propto-\log \exp{\left[ -\sum_{i=1}^{n}  \mathcal{L} \left(f_{\bm{w}}(\bm{x}_i), \bm{y}_i \right)\right]} - \log \exp\left[-\frac{\bm{w}^T\bm{w}}{2 \lambda^{-1}}\right] \\
    &= \sum_{i=1}^{n}  \mathcal{L} \left(f_{\bm{w}}(\bm{x}_i), \bm{y}_i \right) + \frac{1}{2}\lambda \left\Vert\bm{w}\right\Vert_2^2,
\end{split}
\end{equation}
{where $\lambda$ is the inverse of the Gaussian prior variance. Note that the energy function $U$ is equivalent to the loss function in normal neural network training with SGD momentum, and the Gaussian prior with variance $\lambda^{-1}$ reduces to a commonly used $L_2$ regularization term in the energy (loss) function. In practice, we choose $\lambda=3\times10^{-5}$ which forms a relatively weak prior assumption because of the high prior variance. We will also research the effect of varying $\lambda$.}

\subsubsection{The Cold Posterior with Tempering}
For medical image segmentation, data augmentation proved to be a highly practical and effective strategy to overcome the data scarcity problem~\citep{isensee2021nnu,campello2021multi,0A}. Extensive data augmentation is also explicitly performed in the nnU-Net~\citep{isensee2021nnu}. Data augmentation, however, violates the independent and identically distributed (i.i.d.) assumption of data samples, leading to the so-called \textit{dirty} likelihood~\citep{nabarro2021data}. We proposed to mitigate the undesirable effect of data augmentation on likelihood estimation by tempering~\eqref{eq:likihood}, as recently suggested by Nabarro \textit{et al.}~\citep{wenzel2020good,nabarro2021data}: 
\begin{equation}
    p_{\mathrm{cold}}\left(\bm{w}|\mathcal{D}\right) \propto \exp (-U(\bm{w}) / T),
    \label{eq:cold_post}
\end{equation}
where $T$ is named ``temperature'' in analogy to Maxwell-Boltzmann Statistics in physics, to counteract the dirty likelihood effect of data augmentation. A ``cold'' temperature $T<1$ compensates for the increased number of training samples from data augmentation and limits the variance of posterior samples~\citep{nabarro2021data,wenzel2020good}.

\subsection{Bayesian Inference and Voxel-wise Uncertainty}

\label{sec:bayesian_inference}
With the posterior estimation of the weight distribution $p(\bm{w}|\mathcal{D})$, the prediction on a test image $\bm{x}^{*}$ can be made by integration: 
\begin{equation}
    p(\bm{y}^{*}|\bm{x}^{*}, \mathcal{D})=\int_{\bm{w}} p\left(\bm{w}|\mathcal{D}\right) p\left(\bm{y}^{*}|\bm{x}^{*}, \bm{w}\right) d\bm{w},
    \label{eq:marginalization}
\end{equation}
which, however, cannot be solved analytically without restrictive assumptions on the exact form of likelihood and prior model (e.g., Gaussian)~\citep{gal2016dropout, blei2017variational}. We approximate the integration in~\eqref{eq:marginalization} using the Monte-Carlo method, which is assumption-free and scalable to network sizes:
\begin{equation}
    p(\bm{y}^{*}|\bm{x}^{*}, \mathcal{D}) \approx \frac{1}{M} \sum_{j=1}^{M} p\left(\bm{y}^{*}|\bm{x}^{*}, \bm{w}_j\right),
    \label{eq:monte_carlo_integration}
\end{equation}
where $\{\bm{w}_j\}_{j=1}^{M}$ are the $M$ samples drawn from the posterior distribution $p\left(\bm{w}|\mathcal{D}\right)$ and $p\left(\bm{y}^{*}|\bm{x}^{*}, \bm{w}_j\right)=f_{\bm{w}_j}(\bm{x}^{*})$ is the voxel-wise probabilistic prediction made by the network with weight $\bm{w}_j$. {In practice, the $M$ samples are saved checkpoints during the training or posterior sampling process and these samples can form an ensemble without the need to train multiple ensembles.} Subsequently, we can estimate the predictive uncertainty map based on the voxel-wise binary entropy $H_c$ of each class or the categorical distribution {entropy $\mathcal{H}$}~\citep{mehrtash2020confidence}:
\begin{align}
    \mathcal{H}_c &= -p_c  \log p_c - (1-p_c) \log (1-p_c),\\
    \mathcal{H} &= -\sum_{c=1}^{C} p_c  \log p_c,
\end{align}
where $p_c = p(\bm{y}_{i}^{*}=c|\bm{x}^{*}, \mathcal{D})$.

\subsection{Posterior Sampling via Hamiltonian Monte Carlo}
\label{sec:post_sampling_hmc}
HMC is an MCMC variant that can effectively generate samples conforming to a given distribution~\citep{neal2011mcmc, chen2014stochastic}, scalable to high dimensionality~\citep{speagle2019conceptual}. In this section, we introduce the HMC sampling of the CP distribution defined in~\eqref{eq:cold_post}.
\subsubsection{Stochastic Gradient Hamiltonian Monte Carlo}

\begin{figure}[htb!]
    \centering
    \includegraphics[width=0.8\textwidth]{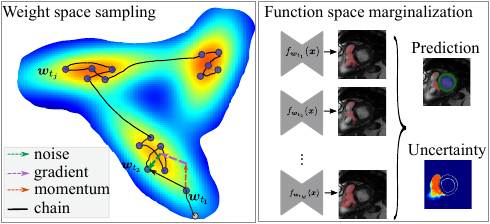}
    \caption{With a limited amount of training data, the network admits infinite weight solutions that can explain the training set. The posterior of weights models the probability density of the solution space, which is characterized by multiple local optima. The HMC chain (black line) is guided by the momentum (red arrow) which accumulates the gradient (purple arrow) to approach the local optima. The noise (green arrow) encourages the exploration of the low-loss surface. Multiple local optima can be visited by the chain via the annealing strategy. {The weight space sampling is essentially similar to training the networks with SGD with momentum. In practice, checkpoints during the chain simulation are saved as posterior samples to form ensembles for function space marginalization.} }
    \label{fig:teaser}
\end{figure}

We propose drawing samples from the CP~\eqref{eq:cold_post}, by HMC~\citep{neal2011mcmc} sampling. HMC builds a Markov chain by simulating the motion of the particles in an energy field $U(\bm{w})$ at position $\bm{w}$ with momentum $\bm{r}$. The particle dynamics is governed by the Hamiltonian, $H(\bm{w}, \bm{r})=U(\bm{w})+\frac{1}{2}\bm{r}^{T}\bm{M}^{-1}\bm{r}$, in which the potential energy $\nabla U(\bm{w})$ drives the particle to a low energy state (equivalently, the low loss region in the weight space). The auxiliary momentum term $\frac{1}{2}\bm{r}^{T}\bm{M}^{-1}\bm{r}$ simulates the kinetic energy that makes the particle explore the low-energy surface. Without loss of generality, the mass $\bm{M}$ can be set as the identity matrix. 

The HMC dynamics simulation requires the evaluation of the full batch gradient $\nabla U(\bm{w})$~\citep{neal2011mcmc}. In practice, however, we only have access to the stochastic gradient estimated on a mini-batch of size $n_b$:
\begin{equation}
\begin{split}
      \nabla \Tilde{U}(\bm{w}) &= \frac{1}{n_b}\sum_{i=1}^{n_b} \left[ \nabla \mathcal{L} \left(f_{\bm{w}}(\bm{x}_i), \bm{y}_i \right) + \lambda \bm{w}\right]   \\
        &= \frac{1}{n} \left[\nabla {U}(\bm{w}) + \mathcal{N}(0, 2\bm{V}) \right],
\end{split}
\end{equation} 
where $U$ is defined in \eqref{eq:loss_function} {and $\nabla \frac{1}{2}\Vert\bm{w}\Vert_2^2 = \bm{w}$}. The stochastic gradient estimation $\nabla \Tilde{U}(\bm{w})$ introduces additional noise of strength $\bm{V}$ to the true gradient. In the presence of such noise, the stationary distribution of the HMC samples is no longer the target distribution~\citep{chen2014stochastic}. 

To address this problem, we propose to use the Stochastic Gradient Hamiltonian Monte Carlo~(SGHMC)~\citep{chen2014stochastic}, which introduces a friction term that compensates for the stochastic gradient noise and a Gaussian noise to the momentum update such that the dampening friction matches the noise level. {In practice, the tempered posterior by $T$ in~{\eqref{eq:cold_post}} leads to the Hamiltonian $H(\bm{w}, \bm{r})= \frac{1}{T}U(\bm{w})+\frac{1}{2}\bm{r}^{T}\bm{M}^{-1}\bm{r}$, and the gradient scales linearly to $\frac{1}{T}\nabla \Tilde{U}(\bm{w})$.} The Markov chain can be simulated according to the discrete form of SGHMC: 
\begin{equation}
    \left\{
    \begin{aligned}
        \bm{w}_{t+1} &= \bm{w}_{t} + \bm{r}_t \\
        \bm{r}_{t+1} &= (1-\mu)\bm{r}_{t} - \frac{1}{T}\eta_t{n} \nabla \Tilde{U}(\bm{w}_t) + \sqrt{{2\eta_t\mu}}\mathcal{N}(0, \mathbb{I}),
    \end{aligned}
    \right.
    \label{eq:sghmc_discrete}
\end{equation}
where $\mu$ is the friction coefficient, $\eta_t$ is the step size of HMC simulation. {Note that the momentum update rule in~{\eqref{eq:sghmc_discrete}} is equivalent to the following form:} 
\begin{equation}
    \bm{r}'_{t+1} = (1-\mu)\bm{r}'_{t} - \eta_t{n} \nabla \Tilde{U}(\bm{w}_t) + \sqrt{{2\eta_t\mu}T}\mathcal{N}(0, \mathbb{I})
\end{equation}
{by multiplying $T$ on both sides and use $\bm{r}'=T\bm{r}$ to replace the original $\bm{r}$}. The dynamics of $(\bm{w}, \bm{r})$ in \eqref{eq:sghmc_discrete} yield samples whose stationary distribution is exactly $p_{\mathrm{cold}}(\bm{w}|\mathcal{D})$. This can be strictly proven via the Fokker-Planck-Equation of the stationary distribution of SGHMC~\citep{sarkka2019applied, chen2014stochastic}. 

Here, we note that when $T=0$, {\eqref{eq:sghmc_discrete}} is exactly the update rule of SGD with momentum, where $1-\mu$ is equivalent to the momentum term and $\eta_t$ is the learning rate. The length of the Markov chain is the number of iterations in network training using SGD-momentum. In this case, the single source of the stochastic noise is the gradient estimation noise in $\Tilde{U}(\bm{w}_t)$ of strength $\bm{V}$. As $T$ increases, additional noise is injected and perturbs the gradient direction, which can be considered as an SGD-momentum process with elevated gradient estimation noise. {In summary, the HMC sampling process is equivalent to the network optimization process as is shown in~{\eqref{eq:sghmc_discrete}} and, thus, comes at no additional cost. The optimizer performs as a posterior sampler and the sampling process is in essence saving the checkpoints.} The overview of the proposed method is shown in Fig.~{\ref{fig:teaser}}: the gradient (potential force) drives the chain to high-posterior-density regions, and the momentum term and the injected noise keep the chain exploring the vicinity of a local optimum.

\subsubsection{Annealing Learning Rate and Sample Thinning}

The learning rate controls the convergence in optimization and Monte Carlo sampling. We present a novel approach to reschedule the learning rate for more accurate posterior sampling. Specifically, to let the chain explore a broader area and prevent it from converging to a single point, we set the learning rate as a constant non-zero value after the $\gamma$ fraction of the training budget. To capture the multimodal posterior geometry that is typical of complex DNNs~\citep{zhang2019cyclical, huang2017snapshot}, we further propose to use cyclical annealing training such that the Markov chain can visit multiple modes of the posterior. We divide the training budget of $T_E$ epochs into $N_C$ cycles and each cycle consumes $T_c = \frac{T_E}{N_C}$ epochs. In particular, we propose to use a high learning rate at the beginning of each cycle such that the perturbation is strong enough to drive the chain into various posterior modes. 

The overall learning rate for the Hamiltonian dynamics simulation is formulated as:
\begin{equation}
\label{eq:cyclical_lr}
    \eta_{t_e} = 
    \begin{cases}
        \eta_r&, t_c < T_r \\
        \eta_0\cdot \left(1 - \frac{\min\left\{t_c, \gamma T_c \right\}}{T_c} \right)^{0.9} &, t_c \geq T_r.
    \end{cases}
\end{equation}
where $t_c = t_e\ \mathrm{mod}\ T_c$ is the intra-cycle epoch number and $\eta_r$ is the high restart learning rate which was set for the first $T_r$ epochs in each cycle.

The first $\gamma$ fraction of training epochs are considered as the burn-in stage of SGHMC~\citep{zhang2019cyclical}. The weights computed at each iteration after the burn-in stage can be seen as a sample of the posterior distribution. However, a single iteration causes a marginal change in weights in the Markov chain, and the consecutive samples can be highly correlated. The auto-correlation between samples significantly reduces the number of effective samples in a Markov chain~\citep{hammersley2013monte}. Moreover, collecting all samples after the burn-in stage requires substantial disk space while the inference would be extremely time-consuming. After the burn-in stage, we, therefore, adopt the sample thinning strategy~\citep{hammersley2013monte} to only collect samples at the end of every fourth epoch (every 1000 iterations): $
    \bm{W} = \left\{ \bm{w}_{t_j} | t_j\ \mathrm{mod} \ T_c \geq \gamma, 1 \leq t_j \leq T_E, t_j \ \mathrm{mod} \ 4 =0 \right\}$.

\subsection{Weight and Functional Space Diversity}
\label{sec:func_div}

We differentiate two types of diversity for weights $\bm{w}$ and function $f_{\bm{w}}$, respectively. Although both are relevant to uncertainty estimation, their relationship is complex and is largely understudied. To investigate the diversity of weights, we use mutual cosine similarity~\citep{larrazabal2021orthogonal} as a metric for weight space diversity, which is defined as:
\begin{equation}
    \mathrm{sim}_{\mathrm{cos}}(\bm{w}_i, \bm{w}_j) = \frac{<\bm{w}_i, \bm{w}_j>}{\Vert\bm{w}_i\Vert_2\cdot\Vert\bm{w}_j\Vert_2}.
    \label{eq:cos_sim_def}
\end{equation}
Additionally, we monitor the volume of the high-dimensional space that the chain explored, via rectangular approximation:
\begin{equation}
    \mathrm{vol}(\bm{W})= \prod_{s=1}^{n_\sigma}\sigma_s,
    \label{eq:vol_def}
\end{equation}
where $\sigma_s$'s are the first singular values of the matrix $\bm{W}= [\bm{w}_1, \bm{w}_2, \dots, \bm{w}_M]$. {In practice, we choose $n_\sigma=5$ because the first five can explain at least $90\%$ of the total weight variance.  Additionally, $n_\sigma = 5$ is a computationally practical choice because of the extremely high dimension of weights $\bm{W}$.}

To investigate the diversity of functional space, we propose to evaluate the variation of predictions by $f_{\bm{w}}$ on the validation set. Given two functions $f_{\bm{w}_i}$ and $f_{\bm{w}_j}$, we measure the functional space distance on a validation set $\mathcal{D}_{val}$ as:
\begin{equation}
\begin{array}{l}
     d(f_{\bm{w}_i}, f_{\bm{w}_j}) = 1 - \frac{1}{\vert \mathcal{D}_{val} \vert} \sum\limits_{\bm{x}\in \mathcal{D}_{val}} \mathrm{DSC}(\bm{e}\circ f_{\bm{w}_i}(\bm{x}), \bm{e} \circ f_{\bm{w}_j}(\bm{x})),
\end{array}
    \label{eq:diversity_measure}
\end{equation}
where $\bm{e}=\mathbb{I}\left(f_E(\bm{x})\neq \bm{y}\right)$ indicates where the ensemble prediction $f_E(\bm{x})$ make erroneous predictions compared to the ground truth $\bm{y}$ at voxel-level. We note that focusing only on such misclassified voxels can better manifest the difference between functions, because in practice, most of the voxels in an image $\bm{x}$ are correctly classified (e.g., background), leading to an over-optimistically high agreement despite the diversity in organ segmentation. 

\subsection{Voxel-wise Calibration Metrics}
To quantify the performance of voxel-wise calibration and uncertainty estimation, we use the Expected Calibration Error (ECE)~\citep{guo2017calibration}, the Brier score (Br)~\citep{brier1950verification} and the negative log-likelihood (NLL)~\citep{ovadia2019can}. For a segmentation task with $N$ voxels in total, the confidence score ranging from $0\%$ to $100\%$ is equally divided into $B$ bins and the ECE score is defined as:
\begin{equation}
    \mathrm{ECE} = \sum_{i=1}^{B}  \frac{\vert B_i \vert}{N}  \cdot \left\vert \mathrm{conf}(B_i) - \mathrm{acc} (B_i) \right\vert
\end{equation}
where $B_i$ is the set of voxels whose confidence falls into the $i^{th}$ bin, $\mathrm{conf}(B_i)$ is the mean confidence and $\mathrm{acc} (B_i)$ is the mean accuracy. The Brier score quantifies the deviation of predictive categorical distribution from the ground truth one-hot label:
\begin{equation}
    \mathrm{Br} = \frac{1}{N} \sum_{i=1}^{N} \sum_{c=1}^{C} \left[p(\bm{y}_{i}^{*}=c|\bm{x}^{*}) - (\bm{y}_{i} ==c)\right]^{2}
\end{equation}
and the NLL metric is defined as:
\begin{equation}
    \mathrm{NLL} = -\frac{1}{N} \sum_{i=1}^{N} \sum_{c=1}^{C}  (\bm{y}_{i} ==c) \cdot \log p(\bm{y}_{i}^{*}=c|\bm{x}^{*}).
\end{equation}

\subsection{Image-level Confidence Score and Failure Detection}

\begin{figure}
    \centering
    \includegraphics[width=1\textwidth]{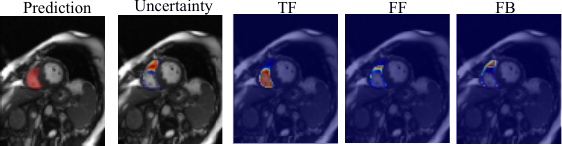}
    \caption{Uncertainty maps indicate possible over- and under-segmentation. We estimate {the true foreground (TF), false foreground(FF), and false background (FB)} using the estimated uncertainty and aggregate them into the final image-level score.}
    \label{fig:compute_score}
\end{figure}

Based on the estimated entropy map, we aggregate the voxel-wise uncertainty and derive an image-level confidence score as a segmentation failure indicator.  {Specifically, we estimate the correct segmentation (true foreground, TF), the over-segmentation (false foreground, FF), and under-segmentation (false background, FB) areas as follows}:
\begin{align}
    \mathrm{TF} &= S_c \cdot (1 - H_c), \\ \nonumber
    \mathrm{FF} &= S_c \cdot H_c, \\ \nonumber
    \mathrm{FB} &= (1 - S_c) \cdot H_c,
\end{align}
where $S_c$ is the segmentation map for class $c$ and $H_c$ is the corresponding entropy map. The final confidence score of the generated segmentation map $S_c$ with uncertainty $H_c$ is given by simulating the Dice coefficient:
\begin{align}
    C(S_c) = \frac{2\vert TF\vert}{2\vert TF \vert + \vert FF \vert + \vert FB \vert}.
\end{align}
Examples of the estimated TF, FF, and FB maps are shown in Fig.~\ref{fig:compute_score}. We detect possible failures based on the computed confidence score $C(S_c)$ and measure the performance of image-level failure detection by the area under the receiver operating characteristic curve (AUC). Empirically, for cardiac MRI applications, a segmentation prediction on a 2D image slice with a dice score lower than 80\% and an average symmetric surface distance (ASSD) greater than 2mm is considered a segmentation failure. 

\section{Experiments}
\subsection{Dataset}
\begin{figure}[h!]
    \centering
    \includegraphics[width=0.96\textwidth]{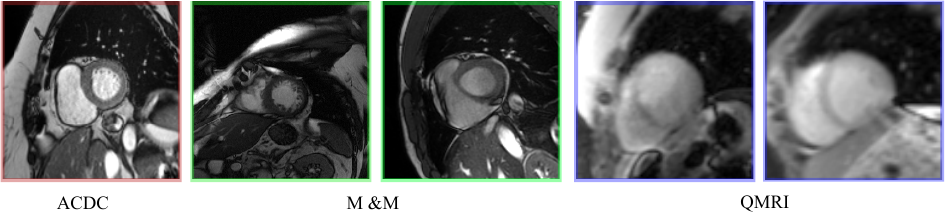}
    \caption{Representative images of ACDC, M\&M and QMRI datasets. ACDC and M\&M are SSFP cine images and the contrast variation is relatively minor. QMRI baseline images have a larger contrast change compared to the training set (ACDC).}
    \label{fig:dataset_vis}
\end{figure}
We evaluated the proposed HMC-CP method on nnU-Net~\citep{isensee2021nnu}, an established U-Net architecture, for the cardiac MRI task. We use the \textbf{ACDC} dataset~\citep{bernard2018deep} for training and validation, which consists of short-axis end-diastolic (ED) and end-systolic (ES) cardiac MRI steady-state free precession (SSFP) cine images of 100 subjects acquired at 1.5T (Aera, Siemens Healthineers, Erlangen, Germany) and 3T (Tim Trio, Siemens Healthineers, Erlangen Germany). {The original training part of the dataset with 100 subjects was randomly split into a training set ($80\%$) and a validation set ($20\%$). The validation set was used for selecting the best temperature and studying the influence of the number of samples on the segmentation and calibration performance. Based on the validation results, we evaluate our methods on the ACDC test set consisting of 50 subjects to test the in-domain performance.} For out-of-domain performance, we tested the proposed method and other methods in comparison on a completely independent dataset, the Multi-center Multi-vendor (\textbf{M\&M}) cardiac MRI~\citep{campello2021multi}, which contains 320 SSFP cine scans of end-systolic (ES) and end-diastolic (ED) images collected from 6 medical centers using different 1.5T -(Siemens Avanto, Germany, Philips Achieva, Netherlands; GE Signa Excite; and Cannon Vantage Orian); and 3T scanners (Siemens Skyra, Germany). Additionally, we also evaluate the proposed method on a quantitative MRI (\textbf{QMRI}) dataset, containing 112 modified look-locker inversion recovery (MOLLI) $T_1$-mapping~\citep{messroghli2004modified} images and $T_2$-prep-based $T_2$-mapping~\citep{giri2009t2} images. The images are collected from 8 healthy subjects at 3T (Prisma, Siemens Healthineers, Erlangen, Germany). Each image contains several (8 for $T_1$-mapping and 5 for T2-mapping) baseline images that were read out during the $T_1$ or $T_2$ relaxation processes. We show some exemplar images of the three datasets in Fig.~\ref{fig:dataset_vis}. Three classes of ground truth labels are provided in the ACDC and M\&M datasets: left ventricle cavity (LV), myocardium (MYO), and right ventricle (RV). For the QMRI dataset, the LV and MYO regions were manually annotated on the second baseline image which has relatively good contrast. 

\subsection{Methods in Comparison}
We implemented several baseline methods including PHi-Seg~\citep{baumgartner2019phiseg}, Bayesian SegNet with Dropout~\citep{kendall2015bayesian}, Deep Ensembles~\citep{lakshminarayanan2016simple, mehrtash2020confidence}, and compared them to the proposed method in terms of both segmentation and uncertainty estimation. We used the automatically configured nnU-Net ~\citep{isensee2021nnu} architecture, which is a commonly used reference for medical image segmentation. {All methods were trained with $1000$ epochs}. We set the initial learning rate to be $\eta_0=0.02$ and used a fixed batch size of $40$ for all methods.

\begin{itemize}
    \item  \textbf{PHi-Seg} We implemented the PHi-Seg~\citep{baumgartner2019phiseg} in the nnU-Net framework with $6$ resolution levels, $5$ latent levels and latent feature depth $4$ for the prior, likelihood and posterior networks. At inference time, we drew $M=30$ realizations of the hierarchical latent features from the prior network output and decoded the features with the likelihood network.
    \item \textbf{MC-Dropout} Following the Bayesian SegNet work~\citep{kendall2015bayesian}, we inserted dropout layers into the innermost three layers on both the encoder- and decoder-side of the U-Net. The dropout rate was set as $p=0.5$ at both the training and testing phases {and we ran $M=30$ forward passes at test time}.  
    \item \textbf{Deep Ensembles} Deep Ensembles of 15 models were trained by SGD-momentum with random initialization for $1000$ epochs with the standard exponential learning rate decay. 
    \item \textbf{SGHMC Variants}\label{sec:method_sghmc} We ran the proposed SGHMC method for $N_C=3$ cycles of $333$ epochs. In each cycle, the first $\gamma=0.60$ fraction of each cycle was the burn-in stage. Afterward, the noise was injected into the momentum update. The noise level is controlled by the temperature $T$ as in \eqref{eq:sghmc_discrete}. To investigate the effect of cold posterior, we trained networks using SGHMC with temperature $T\in\{0, 10^{-6}, 10^{-5}, 10^{-4}, 10^{-3}\}$, where $T=0$ corresponds to SGD-momentum with constant learning rate (SGD-Const)~\citep{zhao2022efficient}. The restart learning rate was set to $\eta_r=0.2$ and restarting epochs $T_r=10$. The checkpoints were collected every $4$ epoch after the burn-in stage as posterior samples until the end of training. The weight evolution in the last training cycle was recorded for the single-mode sampling baselines (SGHMC-Single), and the checkpoints across all three modes form a set of multi-modal weight samples (SGHMC-Multi).   
\end{itemize}

\section{Results}
\subsection{Posterior Geometry and Chain Trajectory of SGHMC}
\label{sec:trajectory}

\begin{figure}[t]
    \centering
    \includegraphics[width=1\textwidth]{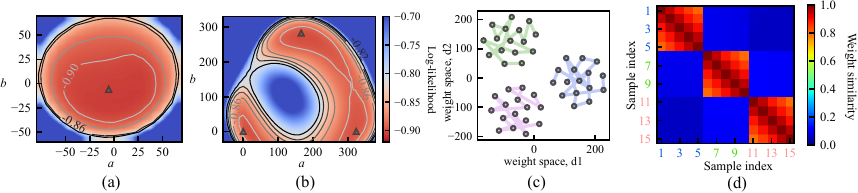}
    \caption{Loss landscape and chain trajectory during training: (a) The loss landscape around a MAP solution. (b) Applying cyclical training promotes the diversity of solutions. The triangular marks indicate the three modes of solutions on the loss surface in three training cycles\protect\footnotemark[1]. (c) The t-SNE map of the collected weight samples illustrates three clusters of local weight samples. (d) Cosine similarity of weight samples collected in three cycles, suggests that weights drawn from a single cycle (mode) of the chain correlate with each other, while weight modes from different cycles are diverse.}
    \label{fig:loss_landscape_trajectory}
\end{figure}
\footnotetext[1]{We visualize the training trajectory of the loss landscape of trained U-Nets on 2D planes. For the checkpoints belonging to the same posterior mode $\bm{W}=\{\bm{w}_{t_j}\}_{j=1}^M$, we perform singular value decomposition on centered $\bm{W}$ to find the first five principal components $\bm{v}_p, p\in\{1, 2, \dots, 5\}$. The validation loss is then visualized via evaluating $\mathcal{L}(\bar{\bm{w}}+a\bm{v}_2+b\bm{v}_3)$ as a function of $(a, b)$, where $\bar{\bm{w}}=\frac{1}{M}\sum_{j=1}^{M}\bm{w}_{t_j}$. For weights $\bm{w}_1$, $\bm{w}_2$ and $\bm{w}_3$ drawn from three posterior modes, we performed the Gram-Schmidt orthogonalization~\citep{garipov2018loss} of $\bm{w}_2-\bm{w}_1$ and $\bm{w}_3-\bm{w}_1$ and used the resultant orthogonal vectors $\bm{u}$ and $\bm{v}$ as the base. We visualize the surface of $\mathcal{L}(\bm{w}_1+a\bm{u}+b\bm{v})$ with varying $(a, b)$~\citep{garipov2018loss}.}

Fig.~\ref{fig:loss_landscape_trajectory} (b) depicts the loss surface on the interpolated plane collected from three training cycles, illustrating the multi-modality of U-Net solution space. Via cyclical learning, the weight iterations visited multiple posterior modes, which can be observed from the t-SNE visualization of the training trajectory in Fig.~\ref{fig:loss_landscape_trajectory} (c). Fig.~\ref{fig:loss_landscape_trajectory} (d) visualizes the cosine similarity~\citep{larrazabal2021orthogonal} of checkpoints collected in three cycles which shows that the local weight checkpoints are similar to each other, while the cyclical training promotes the orthogonality of weights in different modes.

\subsection{Diversity in Function Space}
\label{sec:div_func_res}

\begin{figure}[htb!]
    \centering
    \includegraphics[width=1\textwidth]{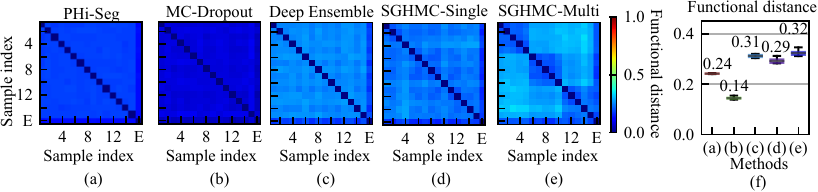}
    \caption{Confusion matrices that show the diversity of functions of PHi-Seg (a), MC-Dropout (b), Deep Ensembles (c), and our proposed SGHMC variants, SGHMC-Single (d) and SGHMC-Multi (e). The ensemble of all function instances is denoted as E, at the lower-right corner of the matrices. Each entry in the confusion matrix represents the mutual distance in the function space of two functions, defined in Sec.~\ref{sec:func_div}. (f) sums up the functional distance values from (a) to (e) and illustrates the mean of rows in the confusion matrices.}
    \label{fig:confusion_mat}
\end{figure}

A high degree of diversity in the function space leads to better uncertainty estimation. We analyzed the functional diversity of all methods in comparison via evaluating the distance between function instances defined in \eqref{eq:diversity_measure}. The result is shown in Fig.~\ref{fig:confusion_mat} (a)-(e) as confusion matrices. The mutual diversity levels for all methods are summarised in Fig.~\ref{fig:confusion_mat} (f). We show that the PHi-Seg and MC-Dropout methods have lower functional diversity in the functional space compared to Deep Ensembles and the proposed HMC variants, namely SGHMC-Single and SGHMC-Multi. The SGHMC-Single model yielded slightly lower diversity than that of Deep Ensembles, however, the SGHMC-Multi model showed the highest diversity, surpassing Deep Ensembles.

\subsection{The Effect of Cold Posterior}
\begin{figure}[h]
    \centering
    \includegraphics[width=\textwidth]{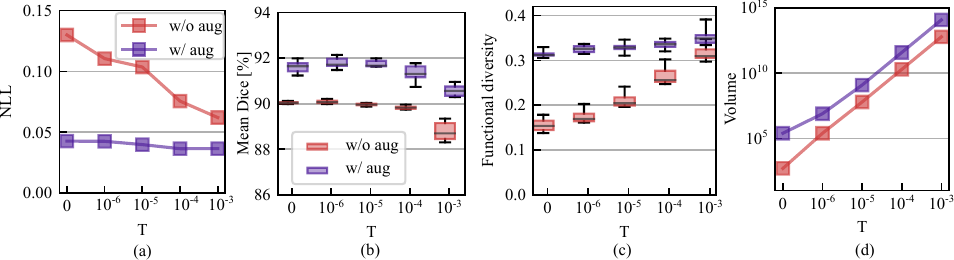}
    \caption{{The cold posterior effect: (a) Calibration performance with various temperatures. (b) Mean Dice over LV/MYO/RV on the validation set of each posterior sample. (c)-(d) The functional and weight space diversity w.r.t. varying temperature.}}
    \label{fig:cold_posterior}
\end{figure}
{ We also studied the effect of varying temperatures on the calibration and segmentation in both with or without augmentation cases. Fig.~{\ref{fig:cold_posterior}} (a) shows the calibration performance variation with an increasing temperature from $T=0$ to $10^{-3}$. From the figure, we observe that the model is not in favor of a cold posterior when augmentation is turned off, as the NLL consistently improves as the temperature increases from $0$ to $10^{-3}$. However, the NLL is relatively less sensitive to the changes in temperature when data augmentation is on and the NLL drop is relatively marginal compared with the without augmentation case. Fig.~{~\ref{fig:cold_posterior}} (b) shows that the mean Dice across LV/MYO/RV drops in both cases as the temperature increases. We conjecture that this is because of the sampling on sub-optimal loss levels because of injected noise. The best segmentation accuracy is achieved at $T=10^{-5}$.}  {Fig. {~\ref{fig:cold_posterior}} (c) - (d) reveals that higher temperature drives the chain to explore broader weight space because the weight volume increases exponentially with an increasing temperature. However, the functional diversity is more sensitive to the weight space volume change as is shown in Fig.~{\ref{fig:cold_posterior}} (c) when the augmentation is turned off. When the augmentation is on, we observe that a cold posterior at $T=10^{-5}$ provides good calibration and improved segmentation performance.} In the following, we use $T=10^{-5}$ to evaluate our method on the test sets. 

\subsection{The Effect of Prior}
\begin{figure}[h]
    \centering
    \includegraphics{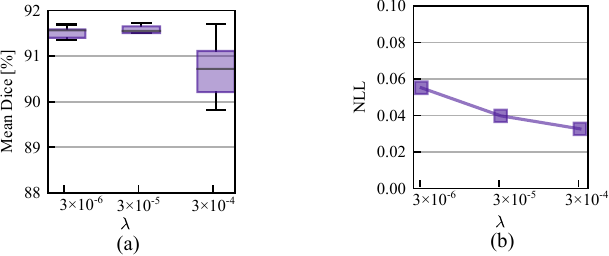}
    \caption{{The effect of prior $\lambda$:  (a) Mean Dice over LV/MYO/RV on the validation set of each posterior sample. (b) Calibration performance with various prior strengths.}}
    \label{fig:prior_sensitiv}
\end{figure}

{In Fig.~{\ref{fig:prior_sensitiv}}, we study the effect of varying prior strength $\lambda$ defined in~{\eqref{eq:loss_function}}. Smaller $\lambda$ indicates a higher prior variance and thus a weaker prior assumption. From Fig.~{\ref{fig:prior_sensitiv}} (a) - (b), we observe that the stronger prior assumptions with $\lambda=3\times10^{-4}$ cause a significant performance drop in Dice, but the calibration performance measured by NLL was improved. As the prior strength increases, its contribution to the posterior geometry is more pronounced and the likelihood part that fits the training data can end up with a sub-optimal level and thus lead to a lower accuracy. However, smaller prior $\lambda=3\times10^{-6}$ can cause a slight performance drop compared with $\lambda=3\times 10^{-5}$, which shows that a proper regularization is beneficial to the accuracy.}

\subsection{Calibration and Segmentation Performance}
\begin{figure*}[h]
    \centering
    \includegraphics[width=\textwidth]{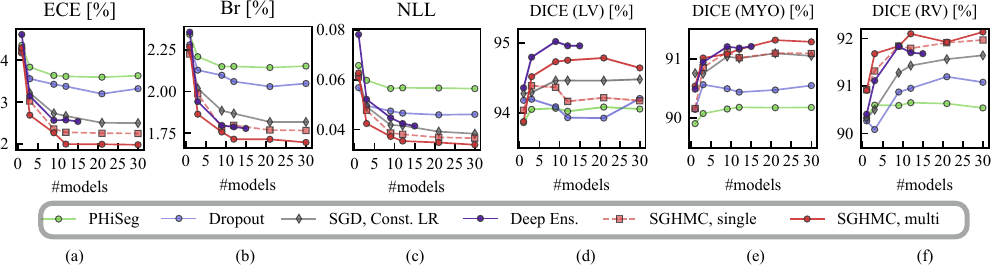}
    \caption{{Calibration quality and segmentation performance as a function of the number of models averaged on the validation set. (a) - (c) illustrate calibration metrics: ECE (a), Br (b), and NLL (c) versus the number of models averaged.} Averaging $M \geq 15$ models using SGHMC significantly improves the network calibration averaged compared to relying solely on a single weight setting. Ensembles of SGHMC samples consistently improve the segmentation and achieve better performance than MC-Dropout and PHi-Seg. The best calibration is achieved by SGHMC with a cyclical annealing learning rate (SGHMC-Multi). (d-f) illustrate the segmentation performance versus several models for three segmentation classes:  LV (d), MYO (e), and RV (f). Ensembles of multi-modal SGHMC samples achieved the best performance on RV and MYO.}
    \label{fig:calib}
\end{figure*}

In Fig.~\ref{fig:calib}, we show the calibration quality and segmentation performance as a function of $M$, which is the number of models (predictions) averaged on the in-distribution validation set. The figure shows that the segmentation performance of the PHi-Seg framework is not competitive with other methods of comparison. Ensembling consistently improves the segmentation performance for all methods, however, the segmentation performance improvement via ensembling MC-Dropout and PHi-Seg predictions is relatively marginal compared to Deep Ensembles and HMC, in accordance with our finding that PHi-Seg and MC-Dropout lack functional diversity. On the in-distribution validation set, an ensemble of $30$ cyclical SGHMC (SGHMC-Multi) samples achieved the best performance. Fig.~\ref{fig:calib} (a)-(c) list the calibration results measured by ECE, Brier score, and NLL respectively. From the figures, we observe that a single model $M=1$ has poor calibration for all methods in comparison while combining more predictions consistently improves the model calibration. MC-Dropout improves calibration but is not as good as the ensemble of SGHMC with a constant learning rate (SGD-Const) at $T=0$. Compared with SGD-Const ($T=0$), SGHMC-Single ($T=10^{-5}$) achieved better calibration performance, which was further surpassed by SGHMC-Multi, which ensembles from multiple posterior modes. 
 
\input{tabs/tab_segmentation_dice}
\begin{figure}
    \centering
    \includegraphics{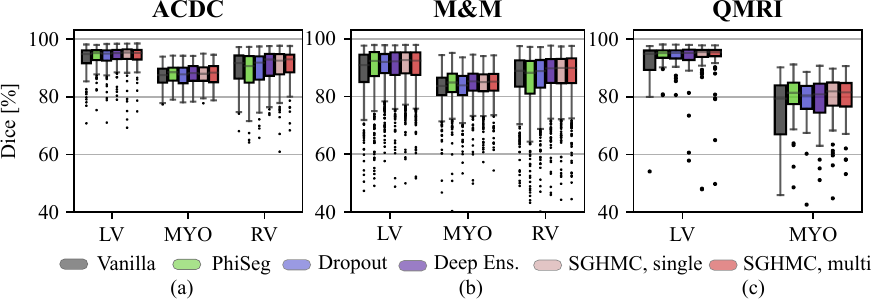}
    \caption{Segmentation performance evaluated on ACDC, M\&M, and QMRI datasets. Both HMC variants (single- and multi-modal samples) substantially improved the segmentation performance compared to the single model prediction (Vanilla), especially on the QMRI dataset with a strong domain shift.}
    \label{fig:dice_performance}
\end{figure}
The segmentation performance measured by Dice score is listed in Fig.~\ref{fig:dice_performance} and Table~\ref{table:segmentation_dice}. The results show that the proposed method improves segmentation performance compared to the Vanilla model on all test sets. This indicates that Bayesian inference via averaging posterior samples leads to more accurate prediction. On the ACDC test set and M\&M, the performance of Deep Ensemble and the proposed method are marginally better than other methods in comparison. There is little difference between the segmentation performance of the single-modal and multi-modal variants of the proposed model. {Additionally, we observe that there exists little difference between the proposed method and traditional methods like Deep Ensembles. However, it is not our primary purpose to significantly increase the segmentation performance against the traditional methods. Instead, we are more interested in the calibration and uncertainty quantification performance.} 

{The results also show that domain shifts cause a performance drop, but our proposed method by marginalizing HMC samples is more robust to the contrast changes.} Comparing the results on the three datasets, we observe a drastic performance drop as the domain shift increases from cine (ACDC, M\&M) to QMRI. The MYO Dice of the vanilla model drops from $87.03\%$ to $63.88\%$ on QMRI. However, on the QMRI dataset with the largest domain shift, the proposed method exhibits high robustness and achieved the highest Dice score on both LV ($89.88\%$) and MYO ($76.20\%$). 

\input{tabs/tab_calibration}
\begin{figure}
    \centering
    \includegraphics[width=\textwidth]{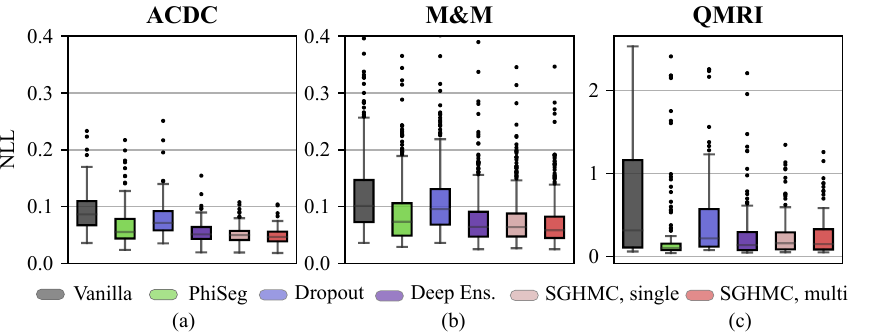}
    \caption{Voxel-wise calibration performance measured by NLL on the ACDC, M\&M, and QMRI datasets. HMC variants significantly improved the calibration and achieved the best calibration score measured by NLL.}
    \label{fig:calib_performance}
\end{figure}
In Fig.~\ref{fig:calib_performance} and Table~\ref{table:calibration}, we report the voxel-wise calibration metrics of methods in comparison. The figure and table show that the calibration performance decreases consistently as the domain shift becomes larger. Overall, the multi-modal SGHMC predictions deliver the best calibration performance, significantly better than baseline methods like MC-Dropout and Phi-Seg on the cine datasets (ACDC and M\&M). This is in accordance with the results in Sec.~\ref{sec:div_func_res} which show that the HMC samples have higher functional diversity. On the QMRI dataset, the PhiSeg network achieved the best ECE and Brier score, but the single-modal and multi-modal SGHMC variants have the lowest NLL. Comparing the results of Dropout and Deep Ensemble, the proposed method is robustly well-calibrated even in the presence of a large domain shift.

\subsection{Automated Failure Detection}
\input{tabs/tab_auc}
\begin{figure}[htb!]
    \centering
    \includegraphics{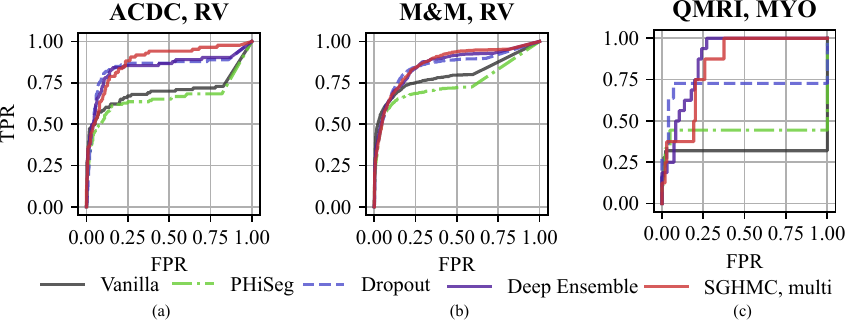}
    \caption{ROC of RV segmentation failure detection on ACDC (a) and M\&M datasets (b), and MYO segmentation failure detection (c) on QMRI dataset.}
    \label{fig:roc_curves}
\end{figure}
With the proposed image-level uncertainty score, we could automatically detect segmentation failure for the test datasets. Table~\ref{table:failure_detection_auc} lists the AUC values of failure detection on three datasets. MC-Dropout, deep ensemble, and the proposed HMC variants achieved similar performances in failure detection of LV and MYO segmentations on the ACDC dataset. However, the proposed method can detect segmentation failure of the most challenging anatomy RV better with an AUC of 88.21\%, which is higher than deep ensemble (84.57\%) and MC-Dropout (84.57\%). This is also reflected in the ROC curve in Fig.~\ref{fig:roc_curves} (a) which shows that the ROC of the proposed method encloses that of Deep Ensemble and MC-Dropout. The improvement is also observed in M\&M and QMRI datasets. The proposed method achieved a remarkable AUC (91.47\%) on the QMRI dataset, significantly outperforming the MC-Dropout(70.75\%). Despite the good voxel-wise calibration, PHi-Seg suffers from severe silent failures on QMRI with an AUC of 43.58\%. More detailed ROC curves are listed in Appendix~\ref{sec:append_eval_res}, Fig.~\ref{fig:roc_full}.

\subsection{Qualitative Results}
\label{sec:qualitative}

 \begin{figure}[h]
     \centering
     \includegraphics{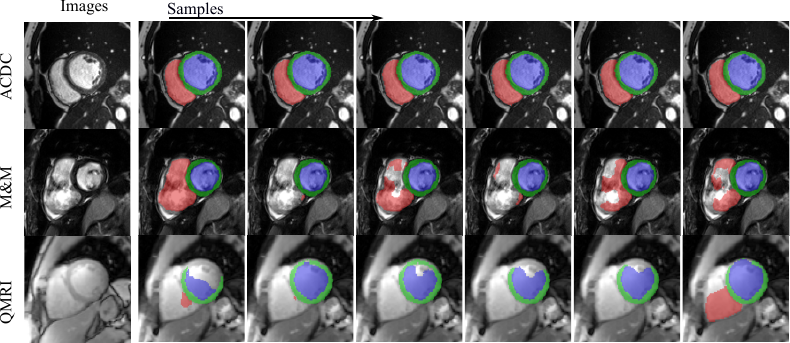}
     \caption{{Visualization of predictions by HMC samples on both in-domain (ACDC) and out-of-domain (M\&M, QMRI) images.}}
     \label{fig:trajectory_vis}
 \end{figure}

{Fig.~{\ref{fig:trajectory_vis}} visualizes the segmentation predictions made by HMC samples. From the figure, we can observe a consistency across these predictions on a certain input, for example, on the cine case from the ACDC dataset. However, on uncertain inputs like the M\&M cine case, the samples tend to make diverse predictions on the RV basal areas. As the contrast change increases further, the network also makes different predictions on the LV blood pool and myocardium because of the low contrast of QMRI images.}
 
\begin{figure}[htb!]
    \centering
    \includegraphics[width=1\textwidth]{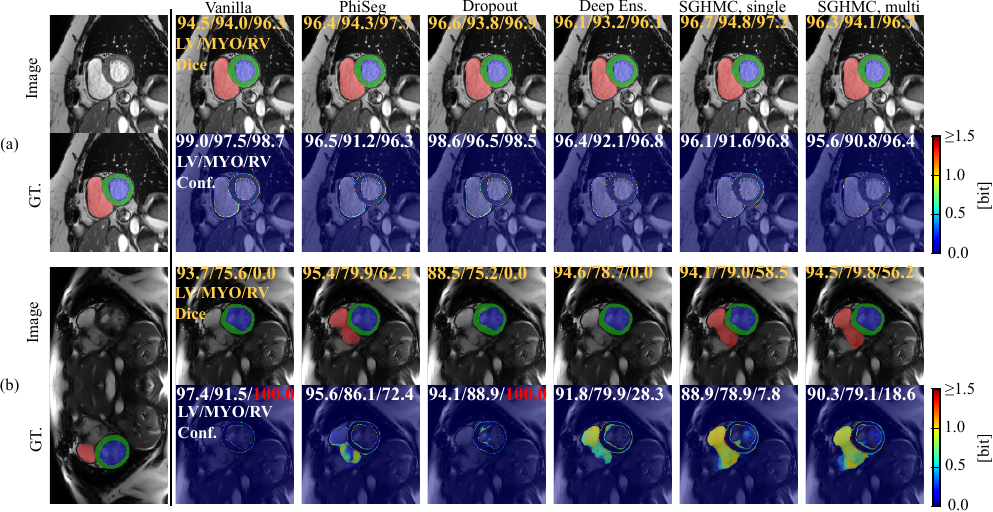}
    \caption{{Segmentation predictions with Dice scores of LV/MYO/RV (the first row) on the QMRI images and the corresponding confidence scores (Conf.) placed on top pf pixel-wise uncertainty maps (the second row).} Red values indicate high confidence scores on segmentation failures.}
    \label{fig:qualitative1}
\end{figure}
\begin{figure}[htb!]
    \centering
    \includegraphics[width=1\textwidth]{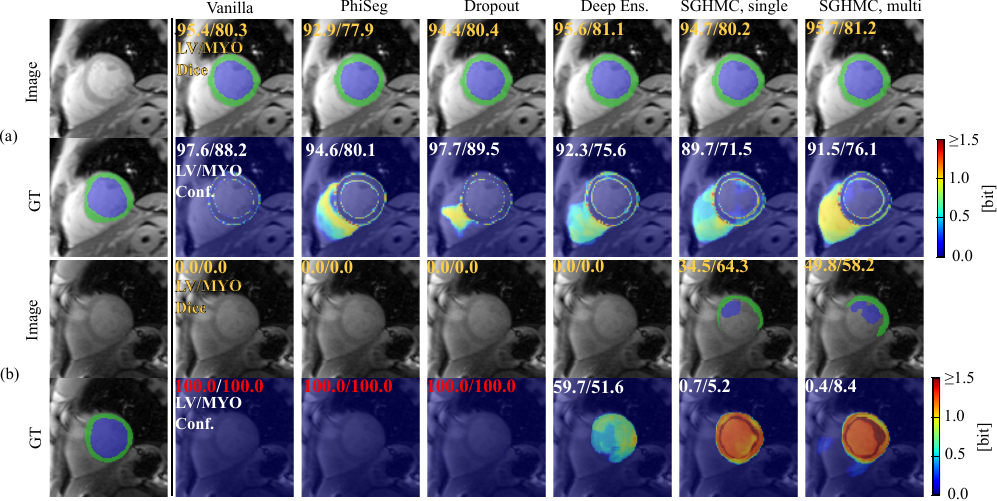}
    \caption{{Segmentation predictions with Dice scores of LV/MYO/RV (the first row) on the QMRI images and the corresponding confidence scores (Conf.) placed on top pf pixel-wise uncertainty maps (the second row).} The Dice values of LV and MYO are listed on the first row and the confidence scores are listed on the second row. Red values indicate high confidence scores on segmentation failures.}
    \label{fig:qualitative2}
\end{figure}
In Fig.~\ref{fig:qualitative1}, we visualize the predictions on the cine images and estimated uncertainty maps produced by all methods in comparison. Fig.~\ref{fig:qualitative1} (a) depicts a middle slice image on which all the methods make accurate predictions and the uncertainty concentrates only on the border between anatomical structures. However, in all four cases, the vanilla network can only output high uncertainty on the border area, even in the presence of erroneous predictions (i.e. silent failure). In general, for cardiac MRI segmentation tasks, uncertainty, and segmentation failure occur more frequently on the right ventricle. A typical failure case is shown in Figure.~\ref{fig:qualitative1} (b), on which the Phi-Seg outputs zero uncertainty on RV voxels.  

Figure.~\ref{fig:qualitative2} (a) shows a QMRI image case on which the network successfully segmented MYO and LV, and a high level of confidence scores are provided by all the methods. Fig.~\ref{fig:qualitative2} (b) depicts the uncertainty estimation results on a hard QMRI image. Under the strong domain shift, PHiSeg and MC-Dropout failed to detect the segmentation failures. In all cases, our proposed HMC variants demonstrated much better uncertainty estimation performance, without the risk of ``silent failures".

\section{Discussion and Conclusion}
DL models have achieved extraordinary performance for medical image segmentation tasks. However, they are generally miscalibrated and can fail silently. This seriously {undermines DL models' trustworthiness} in clinical utilization. In this work, we propose a framework of Bayesian deep learning for uncertainty estimation and failure detection for DL-based medical image segmentation. 

To make the high-dimensional Bayesian segmentation problem computationally tractable, we formulated it as a posterior sampling problem, which can be solved by HMC. {Diversity in the functional space plays a critical role in uncertainty estimation.} We have shown that, in the context of cardiac MRI segmentation, the HMC can effectively produce network samples that make diverse segmentation predictions with a thinning of 4 epochs in our practice. Moreover, ensembling more than $15$ samples in one chain simulation can provide comparable or even superior performance compared to training $15$ deep ensemble models. {Comparing the functional diversity of various methods, we observe that degraded calibration performance strongly correlates with the lack of functional diversity. HMC samples and Deep Ensembles can produce functions of high diversity and their calibration is also superior to other methods including PHi-Seg and MC-Dropout.} 

We also noticed that cold posterior (CP) is crucial to an accurate and well-calibrated prediction for segmentation purposes. Specifically, we observe that HMC at temperature $T=0$ forms an outstanding baseline for model calibration. This method is also successfully validated on the segmentation of shapes with multiple connected components like lymph nodes~{\citep{salahuddin2023head}}. Moreover, the performance can be further improved by increasing to a cold temperature at $T=10^{-5}$. On the other hand, when the data augmentation is turned off, a cold temperature causes a significant drop in model calibration.

{In general, the marginalization of diverse functions is beneficial to the segmentation accuracy. All methods improve the segmentation accuracy by combining predictions made by various network samples. A relatively higher function space diversity improves the segmentation accuracy, comparing HMC to PHi-Seg and MC-Dropout. However, overly high function space diversity could harm the segmentation accuracy. HMC with a higher temperature $T \geq 10^{-4}$ produces an elevated level of functional diversity but the segmentation accuracy drops. In contrast, the calibration consistently improves with an increasing temperature.}

In particular, our proposed method has addressed the two major issues of current Bayesian learning methods for large, over-parameterized neural networks. First, it makes no restrictive assumptions on weight posterior as the VI family of Bayesian methods (including the VI proxy such as MC Dropout). Second, with our proposed cyclical annealing strategy, the method is highly computationally efficient, only consuming the same computation budget as a single-round standard network training. {The resemblance between HMC and SGD with momentum makes posterior sampling as straightforward as saving checkpoints during network training. Nonetheless, we acknowledge that with more computational resources, Deep Ensembles can also be trained in parallel, which is also time-efficient and further enables combining predictions from various network architectures. We did not cover the ensembling of different architectures in our work, which we see as a limitation. However, our method still forms an outstanding baseline with less training effort and energy consumption.}

{In practice, HMC parameters like batch size and learning rate can directly affect the simulation (training) dynamics, because the batch size determines the strength of the gradient estimation noise and the learning rate controls the range of exploration. We used a radical learning rate to promote diversity but the influence of batch size and learning rate on the final performance has not been fully discussed. We see this as a limitation and an interesting future work direction. }

By the proposed image-level confidence score, we can also automatically detect the possible segmentation failure on each image. We showed that the automatic failure detection is highly robust on both in-domain cine images and QMRI images with a strong domain shift with AUC above $86\%$. The HMC approach is especially robust to large domain shifts like cine to QMRI, being the most robust one to detect myocardial segmentation failure in quantitative CMR with an AUC of $91\%$. When integrated into the DL workflow, automatic failure detection potentially improves the trustworthiness of DL models deployed on large-scale clinical studies or in daily clinical practice. 

In conclusion, we have proposed a Bayesian learning framework for uncertainty estimation of medical image segmentation, by Hamiltonian Monte Carlo with cold posterior (HMC-CP). HMC-CP is theoretically grounded, computationally efficient, and scalable to large medical image segmentation networks. Our extensive experiments on both in-domain and out-of-domain data showed that the proposed HMC-CP method results in more reliable uncertainty estimation, as well as more accurate image segmentation, compared with a range of state-of-the-art baselines. Importantly, by reliable uncertainty estimation, our method provides a promising direction for improving the trustworthiness of DL models in clinical applications.


\acks{YZ and QT acknowledge the TU Delft AI Initiative for financial support, and the NVIDIA Applied Research Accelerator Program for granting high-performance computing resources. JT and SW are supported by the NWO Start-up grant STU.019.024, the TU Delft - Erasmus MC Convergence Impulse initiative, the European Union (ERC, Vascular ID, 101078711), and the Netherlands Heart Foundation Dekker Grant. TAT is supported by the British Heart Foundation (IRF FS/19/35/34374). IP and TAT are directly and indirectly supported by the University College London Hospitals NIHR Biomedical Research Centre and Biomedical Research Unit at Barts Hospital respectively.}

%
\ethics{The work follows appropriate ethical standards in conducting research and writing the manuscript, following all applicable laws and regulations regarding the treatment of animals or human subjects.}

\coi{We declare we don't have conflicts of interest.}

\clearpage
\bibliography{sample}


\clearpage
\appendix
\section{Detailed evaluation results}
\label{sec:append_eval_res}


\begin{figure}[htb!]
    \centering
    \includegraphics{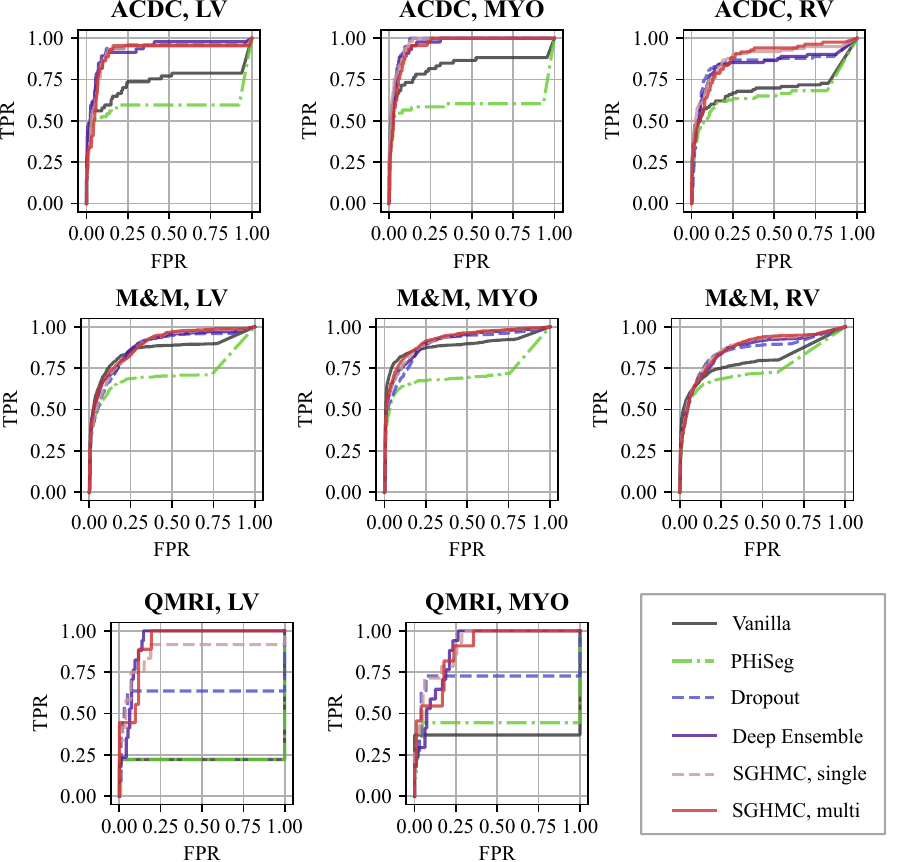}
    \caption{ROC curves of all classes on the ACDC, M\&M and QMRI datasets.}
    \label{fig:roc_full}
\end{figure}


\end{document}

%% file: tabs/tab_segmentation_dice.tex
\begin{table}[h!]
\setlength\extrarowheight{5pt}
\tiny 
\caption{Segmentation performance measured by the Dice score [\%] on ACDC, M\&M, and QMRI datasets. Highlighted are the \best{best} and the \secbest{second best} results.}
\label{table:segmentation_dice}
\centering
\begin{tabular}{P{1.9cm}<{\centering}P{1.1cm}<{\centering}P{1.1cm}<{\centering}P{1.1cm}<{\centering}P{1.1cm}<{\centering}P{1.1cm}<{\centering}P{1.1cm}<{\centering}P{1.2cm}<{\centering}P{1.5cm}<{\centering}P{1.5cm}<{\centering}}
\toprule
\multirow{2}{*}{\textbf{Methods}} & \multicolumn{3}{c}{\textbf{ACDC}} & \multicolumn{3}{c}{\textbf{M\&M}} & \multicolumn{2}{c}{\textbf{QMRI}} \\  \cmidrule(lr){2-4}\cmidrule(lr){5-7}\cmidrule(lr){8-9}
                         & LV           & MYO  & RV  & LV     & MYO     & RV     & LV         & MYO         \\
\midrule
Vanilla\quad\quad {(M=1)}                & 92.80±5.15 &  87.03±3.65 &  90.08±5.71 & 88.64±7.75 &  82.97±5.49 &  86.34±9.36 & 74.17±38.45 &  63.88±31.51 \\
PhiSeg  \quad\quad {(M=30)}                 & 93.64±3.81 &  87.84±3.50 &  88.47±7.32 & 89.79±8.11 &  84.05±5.60 &  84.83±11.16 & \secbest{87.11±25.87} &  \secbest{74.59±21.77}\\
Dropout  {(M=30)}                 & 93.36±4.41 &  87.47±3.66 &  89.50±6.55 & 90.19±6.49 &  83.30±5.45 &  86.45±9.16  & 85.44±28.08 &  72.51±23.96 \\
Deep Ens.   {(M=30)}         & 93.84±3.79 &  87.91±3.54 &  \best{90.92±5.39} & 90.13±6.96 &  \best{84.48±5.25} &  \best{87.43±8.53} & 82.31±31.52 &  70.29±27.04 \\
{SGD Const. LR} { (M=30) }    & 93.70±4.27 & \secbest{87.97±3.57} & 90.61±5.88 & 90.28±6.71 & 84.21±5.14 & 87.26±9.80 & 85.77±27.38 & 73.11±23.80 \\
SGHMC, single { (M=30) }          & 93.54±5.13 &  {87.91±3.45} &  90.16±6.65 & \best{90.48±6.64} &  84.21±5.26 &  \secbest{87.40±8.98} & 86.31±26.11 &  74.22±22.27 \\
SGHMC, multi  {(M=30)}            & \best{93.88±3.82} &  \best{87.97±3.65 }&  \secbest{90.74±5.79} & \secbest{90.31±6.88} &  \secbest{84.43±5.22} &  87.38±9.03 & \best{89.88±20.30} &  \best{76.20±19.16} \\ 
\bottomrule
\end{tabular}
\end{table}

%% file: tabs/tab_calibration.tex
\begin{table}[h!]
\setlength\extrarowheight{5pt}
\tiny 
\caption{Calibration performance measured by the ECE, Br, and NLL on  ACDC, M\&M, and QMRI datasets. Highlighted are the \best{best} and the \secbest{second best} results.}
\label{table:calibration}
\centering
\begin{tabular}{P{1.9cm}<{\centering}P{1.1cm}<{\centering}P{1.1cm}<{\centering}P{1.1cm}<{\centering}P{1.1cm}<{\centering}P{1.1cm}<{\centering}P{1.1cm}<{\centering}P{1.2cm}<{\centering}P{1.1cm}<{\centering}P{1.1cm}<{\centering}}
\toprule
\multirow{2}{*}{\textbf{Methods}} & \multicolumn{3}{c}{\textbf{ACDC}} & \multicolumn{3}{c}{\textbf{M\&M}} & \multicolumn{3}{c}{\textbf{QMRI}} \\  \cmidrule(lr){2-4}\cmidrule(lr){5-7}\cmidrule(lr){8-10}
                         & ECE[\%] $\downarrow$   & Br [\%]$\downarrow$ & NLL[\%] $\downarrow$ & ECE [\%] $\downarrow$ & Br [\%]$\downarrow$ & NLL [\%]$\downarrow$ & ECE [\%]$\downarrow$ & Br [\%]$\downarrow$ & NLL[\%]$ \downarrow$ \\
\midrule
Vanilla  \quad\quad {(M=1)} & 5.45±1.63 & 2.76±0.80 & 9.45±3.87 & 6.49±2.81 & 3.28±1.40 & 11.96±7.25 & 29.66±25.39 & 14.92±12.65 & 76.01±83.30 \\
PhiSeg  \quad\quad {(M=30)}                  & 4.55±1.86 & 2.61±0.96 & 6.91±3.94 & 5.55±2.82 & 3.11±1.44 & 8.68±5.55 & \best{14.87±18.65} & \best{8.03±9.16} & 28.23±48.75\\
Dropout  {(M=30)}            & 5.09±1.71 & 2.69±0.87 & 8.10±3.49 & 6.19±2.50 & 3.23±1.29 & 10.66±5.75  & 22.79±19.06 & 11.60±9.10 & 42.03±47.43 \\
Deep Ens. {(M=15)}                & \secbest{3.49±1.20} & \secbest{2.20±0.63} & 5.55±2.01 & \secbest{4.49±2.14} & \best{2.71±1.12} & 7.51±4.35 & 20.67±19.94 & 10.73±9.33 & 28.61±37.71\\
{SGD Const. LR  (M=30)}       & 3.68±1.40 & 2.27±0.70 & 5.38±1.93 &  4.81±2.35 & 2.81±1.24 & 7.65±4.64 & 19.92±18.65 & 10.35±8.98 & 26.91±33.44 \\
SGHMC, single  {(M=30)}           & 3.54±1.23 & 2.24±0.64 & \secbest{5.17±1.68} & 4.75±2.23 & 2.81±1.19 & \secbest{7.43±4.21} & 20.11±17.11 & 10.33±8.03 & \secbest{26.11±27.29} \\
SGHMC, multi  {(M=30)}            & \best{3.29±1.18} & \best{2.19±0.63} & \best{4.89±1.53} & \best{4.38±2.19} & \secbest{2.72±1.17} & \best{6.91±4.00} & \secbest{19.81±15.60} & \secbest{10.18±7.56} & \best{25.16±24.91} \\ 
\bottomrule
\end{tabular}
\end{table}

%% file: tabs/tab_auc.tex
\begin{table}[h!]
\setlength\extrarowheight{5pt}
\tiny 
\caption{AUC [\%] of failure detection on ACDC, M\&M, and QMRI datasets. Highlighted are the \best{best} and the \secbest{second best} results.}
\vspace{5pt}
\label{table:failure_detection_auc}
\centering
\begin{tabular}{P{2.5cm}<{\centering}P{1.1cm}<{\centering}P{1.1cm}<{\centering}P{1.1cm}<{\centering}P{1.1cm}<{\centering}P{1.1cm}<{\centering}P{1.1cm}<{\centering}P{1.1cm}<{\centering}P{1.1cm}<{\centering}P{1.1cm}<{\centering}}
\toprule
\multirow{2}{*}{\textbf{Methods}} & \multicolumn{3}{c}{\textbf{ACDC}} & \multicolumn{3}{c}{\textbf{M\&M}} & \multicolumn{2}{c}{\textbf{QMRI}} \\  \cmidrule(lr){2-4}\cmidrule(lr){5-7}\cmidrule(lr){8-9}
                         & LV           & MYO  & RV  & LV     & MYO     & RV     & LV         & MYO         \\
\midrule
Vanilla                  & 73.54 & 83.43 & 69.80 & 85.67 & 88.69 & 79.56 & 22.22 & 37.04   \\
PhiSeg                   & 59.41 & 60.05 & 65.88 & 71.02 & 71.31 & 74.10 & 22.01 & 43.58  \\
Dropout                  & \secbest{92.35} & \secbest{96.99} & 84.57 & 87.23 & 89.23 & 84.94 & 62.47 & 70.75  \\
Deep Ens.                & \best{93.21} & 96.42 & 84.57 & \secbest{88.43} & 90.71 & 85.93 & \best{93.68} & \secbest{89.41}  \\
SGHMC, single            & 91.49 & \best{97.27} & \secbest{88.20} & 88.31 & \secbest{91.08} & \best{86.98} & 86.83 & \best{91.47}  \\
SGHMC, multi             & 91.12 & 95.54 & \best{88.21} & \best{89.33} & \best{92.05} & \secbest{86.70} & \secbest{92.88} & 89.20  \\ 
\bottomrule
\end{tabular}
\end{table}

%% file: melba-sample.bbl
\begin{thebibliography}{68}
\providecommand{\natexlab}[1]{#1}
\providecommand{\url}[1]{\texttt{#1}}
\expandafter\ifx\csname urlstyle\endcsname\relax
  \providecommand{\doi}[1]{doi: #1}\else
  \providecommand{\doi}{doi: \begingroup \urlstyle{rm}\Url}\fi

\bibitem[Abdar et~al.(2021)Abdar, Pourpanah, Hussain, Rezazadegan, Liu, Ghavamzadeh, Fieguth, Cao, Khosravi, Acharya, et~al.]{abdar2021review}
Moloud Abdar, Farhad Pourpanah, Sadiq Hussain, Dana Rezazadegan, Li~Liu, Mohammad Ghavamzadeh, Paul Fieguth, Xiaochun Cao, Abbas Khosravi, U~Rajendra Acharya, et~al.
\newblock A review of uncertainty quantification in deep learning: Techniques, applications and challenges.
\newblock \emph{Information Fusion}, 76:\penalty0 243--297, 2021.

\bibitem[Baumgartner et~al.(2019)Baumgartner, Tezcan, Chaitanya, H{\"o}tker, Muehlematter, Schawkat, Becker, Donati, and Konukoglu]{baumgartner2019phiseg}
Christian~F Baumgartner, Kerem~C Tezcan, Krishna Chaitanya, Andreas~M H{\"o}tker, Urs~J Muehlematter, Khoschy Schawkat, Anton~S Becker, Olivio Donati, and Ender Konukoglu.
\newblock Phiseg: Capturing uncertainty in medical image segmentation.
\newblock In \emph{International Conference on Medical Image Computing and Computer-Assisted Intervention}, pages 119--127. Springer, 2019.

\bibitem[Bernard et~al.(2018)Bernard, Lalande, Zotti, Cervenansky, Yang, Heng, Cetin, Lekadir, Camara, Ballester, et~al.]{bernard2018deep}
Olivier Bernard, Alain Lalande, Clement Zotti, Frederick Cervenansky, Xin Yang, Pheng-Ann Heng, Irem Cetin, Karim Lekadir, Oscar Camara, Miguel Angel~Gonzalez Ballester, et~al.
\newblock Deep learning techniques for automatic mri cardiac multi-structures segmentation and diagnosis: is the problem solved?
\newblock \emph{IEEE transactions on medical imaging}, 37\penalty0 (11):\penalty0 2514--2525, 2018.

\bibitem[Blei et~al.(2017)Blei, Kucukelbir, and McAuliffe]{blei2017variational}
David~M Blei, Alp Kucukelbir, and Jon~D McAuliffe.
\newblock Variational inference: A review for statisticians.
\newblock \emph{Journal of the American statistical Association}, 112\penalty0 (518):\penalty0 859--877, 2017.

\bibitem[Blundell et~al.(2015)Blundell, Cornebise, Kavukcuoglu, and Wierstra]{blundell2015weight}
Charles Blundell, Julien Cornebise, Koray Kavukcuoglu, and Daan Wierstra.
\newblock Weight uncertainty in neural network.
\newblock In \emph{International Conference on Machine Learning}, pages 1613--1622. PMLR, 2015.

\bibitem[Brier et~al.(1950)]{brier1950verification}
Glenn~W Brier et~al.
\newblock Verification of forecasts expressed in terms of probability.
\newblock \emph{Monthly weather review}, 78\penalty0 (1):\penalty0 1--3, 1950.

\bibitem[Campello et~al.(2021)Campello, Gkontra, Izquierdo, Mart{\'\i}n-Isla, Sojoudi, Full, Maier-Hein, Zhang, He, Ma, et~al.]{campello2021multi}
V{\'\i}ctor~M Campello, Polyxeni Gkontra, Cristian Izquierdo, Carlos Mart{\'\i}n-Isla, Alireza Sojoudi, Peter~M Full, Klaus Maier-Hein, Yao Zhang, Zhiqiang He, Jun Ma, et~al.
\newblock Multi-centre, multi-vendor and multi-disease cardiac segmentation: the m\&ms challenge.
\newblock \emph{IEEE Transactions on Medical Imaging}, 40\penalty0 (12):\penalty0 3543--3554, 2021.

\bibitem[Carvalho et~al.(2020)Carvalho, Clark, Nicastro, and Kelly]{carvalho2020scalable}
Eduardo~DC Carvalho, Ronald Clark, Andrea Nicastro, and Paul~HJ Kelly.
\newblock Scalable uncertainty for computer vision with functional variational inference.
\newblock In \emph{Proceedings of the IEEE/CVF Conference on Computer Vision and Pattern Recognition}, pages 12003--12013, 2020.

\bibitem[Chen et~al.(2020)Chen, Qin, Qiu, Tarroni, Duan, Bai, and Rueckert]{chen2020deep}
Chen Chen, Chen Qin, Huaqi Qiu, Giacomo Tarroni, Jinming Duan, Wenjia Bai, and Daniel Rueckert.
\newblock Deep learning for cardiac image segmentation: a review.
\newblock \emph{Frontiers in Cardiovascular Medicine}, 7:\penalty0 25, 2020.

\bibitem[Chen et~al.(2014)Chen, Fox, and Guestrin]{chen2014stochastic}
Tianqi Chen, Emily Fox, and Carlos Guestrin.
\newblock Stochastic gradient hamiltonian monte carlo.
\newblock In \emph{International conference on machine learning}, pages 1683--1691. PMLR, 2014.

\bibitem[Chlap et~al.()Chlap, Min, Vandenberg, Dowling, Holloway, and Haworth]{0A}
P.~Chlap, H.~Min, N.~Vandenberg, J.~Dowling, L.~Holloway, and A.~Haworth.
\newblock A review of medical image data augmentation techniques for deep learning applications.
\newblock \emph{Journal of Medical Imaging and Radiation Oncology}.

\bibitem[Czolbe et~al.(2021)Czolbe, Arnavaz, Krause, and Feragen]{czolbe2021segmentation}
Steffen Czolbe, Kasra Arnavaz, Oswin Krause, and Aasa Feragen.
\newblock Is segmentation uncertainty useful?
\newblock In \emph{Information Processing in Medical Imaging: 27th International Conference, IPMI 2021, Virtual Event, June 28--June 30, 2021, Proceedings 27}, pages 715--726. Springer, 2021.

\bibitem[Der~Kiureghian and Ditlevsen(2009)]{der2009aleatory}
Armen Der~Kiureghian and Ove Ditlevsen.
\newblock Aleatory or epistemic? does it matter?
\newblock \emph{Structural safety}, 31\penalty0 (2):\penalty0 105--112, 2009.

\bibitem[Folgoc et~al.(2021)Folgoc, Baltatzis, Desai, Devaraj, Ellis, Manzanera, Nair, Qiu, Schnabel, and Glocker]{folgoc2021mc}
Loic~Le Folgoc, Vasileios Baltatzis, Sujal Desai, Anand Devaraj, Sam Ellis, Octavio E~Martinez Manzanera, Arjun Nair, Huaqi Qiu, Julia Schnabel, and Ben Glocker.
\newblock Is mc dropout bayesian?
\newblock \emph{arXiv preprint arXiv:2110.04286}, 2021.

\bibitem[Fort et~al.(2019)Fort, Hu, and Lakshminarayanan]{fort2019deep}
Stanislav Fort, Huiyi Hu, and Balaji Lakshminarayanan.
\newblock Deep ensembles: A loss landscape perspective.
\newblock \emph{arXiv preprint arXiv:1912.02757}, 2019.

\bibitem[Gal and Ghahramani(2016)]{gal2016dropout}
Yarin Gal and Zoubin Ghahramani.
\newblock Dropout as a bayesian approximation: Representing model uncertainty in deep learning.
\newblock In \emph{international conference on machine learning}, pages 1050--1059. PMLR, 2016.

\bibitem[Gao et~al.(2023)Gao, Zhou, Gao, and Zhuang]{gao2023bayeseg}
Shangqi Gao, Hangqi Zhou, Yibo Gao, and Xiahai Zhuang.
\newblock Bayeseg: Bayesian modeling for medical image segmentation with interpretable generalizability.
\newblock \emph{arXiv preprint arXiv:2303.01710}, 2023.

\bibitem[Garipov et~al.(2018)Garipov, Izmailov, Podoprikhin, Vetrov, and Wilson]{garipov2018loss}
Timur Garipov, Pavel Izmailov, Dmitrii Podoprikhin, Dmitry~P Vetrov, and Andrew~G Wilson.
\newblock Loss surfaces, mode connectivity, and fast ensembling of dnns.
\newblock \emph{Advances in neural information processing systems}, 31, 2018.

\bibitem[Giri et~al.(2009)Giri, Chung, Merchant, Mihai, Rajagopalan, Raman, and Simonetti]{giri2009t2}
Shivraman Giri, Yiu-Cho Chung, Ali Merchant, Georgeta Mihai, Sanjay Rajagopalan, Subha~V Raman, and Orlando~P Simonetti.
\newblock T2 quantification for improved detection of myocardial edema.
\newblock \emph{Journal of cardiovascular magnetic resonance}, 11\penalty0 (1):\penalty0 1--13, 2009.

\bibitem[Gonzalez and Mukhopadhyay(2021)]{gonzalez2021self}
Camila Gonzalez and Anirban Mukhopadhyay.
\newblock Self-supervised out-of-distribution detection for cardiac cmr segmentation.
\newblock In \emph{Medical Imaging with Deep Learning}, 2021.

\bibitem[Gonzalez et~al.(2021)Gonzalez, Gotkowski, Bucher, Fischbach, Kaltenborn, and Mukhopadhyay]{gonzalez2021detecting}
Camila Gonzalez, Karol Gotkowski, Andreas Bucher, Ricarda Fischbach, Isabel Kaltenborn, and Anirban Mukhopadhyay.
\newblock Detecting when pre-trained nnu-net models fail silently for covid-19 lung lesion segmentation.
\newblock In \emph{International Conference on Medical Image Computing and Computer-Assisted Intervention}, pages 304--314. Springer, 2021.

\bibitem[Guo et~al.(2017)Guo, Pleiss, Sun, and Weinberger]{guo2017calibration}
Chuan Guo, Geoff Pleiss, Yu~Sun, and Kilian~Q Weinberger.
\newblock On calibration of modern neural networks.
\newblock In \emph{International Conference on Machine Learning}, pages 1321--1330. PMLR, 2017.

\bibitem[Gustafsson et~al.(2020)Gustafsson, Danelljan, and Schon]{gustafsson2020evaluating}
Fredrik~K Gustafsson, Martin Danelljan, and Thomas~B Schon.
\newblock Evaluating scalable bayesian deep learning methods for robust computer vision.
\newblock In \emph{Proceedings of the IEEE/CVF conference on computer vision and pattern recognition workshops}, pages 318--319, 2020.

\bibitem[Hammam et~al.(2021)Hammam, Ghobadi, Bonarens, and Stiller]{hammam2021real}
Ahmed Hammam, Seyed~Eghbal Ghobadi, Frank Bonarens, and Christoph Stiller.
\newblock Real-time uncertainty estimation based on intermediate layer variational inference.
\newblock In \emph{Proceedings of the 5th ACM Computer Science in Cars Symposium}, pages 1--9, 2021.

\bibitem[Hammersley(2013)]{hammersley2013monte}
John Hammersley.
\newblock \emph{Monte carlo methods}.
\newblock Springer Science \& Business Media, 2013.

\bibitem[Hastings(1970)]{hastings1970monte}
W~Keith Hastings.
\newblock Monte carlo sampling methods using markov chains and their applications.
\newblock 1970.

\bibitem[Hora(1996)]{hora1996aleatory}
Stephen~C Hora.
\newblock Aleatory and epistemic uncertainty in probability elicitation with an example from hazardous waste management.
\newblock \emph{Reliability Engineering \& System Safety}, 54\penalty0 (2-3):\penalty0 217--223, 1996.

\bibitem[Huang et~al.(2017)Huang, Li, Pleiss, Liu, Hopcroft, and Weinberger]{huang2017snapshot}
Gao Huang, Yixuan Li, Geoff Pleiss, Zhuang Liu, John~E Hopcroft, and Kilian~Q Weinberger.
\newblock Snapshot ensembles: Train 1, get m for free.
\newblock \emph{arXiv preprint arXiv:1704.00109}, 2017.

\bibitem[H{\"u}llermeier and Waegeman(2021)]{hullermeier2021aleatoric}
Eyke H{\"u}llermeier and Willem Waegeman.
\newblock Aleatoric and epistemic uncertainty in machine learning: An introduction to concepts and methods.
\newblock \emph{Machine Learning}, 110\penalty0 (3):\penalty0 457--506, 2021.

\bibitem[Isensee et~al.(2021)Isensee, Jaeger, Kohl, Petersen, and Maier-Hein]{isensee2021nnu}
Fabian Isensee, Paul~F Jaeger, Simon~AA Kohl, Jens Petersen, and Klaus~H Maier-Hein.
\newblock nnu-net: a self-configuring method for deep learning-based biomedical image segmentation.
\newblock \emph{Nature methods}, 18\penalty0 (2):\penalty0 203--211, 2021.

\bibitem[Izmailov et~al.(2018)Izmailov, Podoprikhin, Garipov, Vetrov, and Wilson]{izmailov2018averaging}
Pavel Izmailov, Dmitrii Podoprikhin, Timur Garipov, Dmitry Vetrov, and Andrew~Gordon Wilson.
\newblock Averaging weights leads to wider optima and better generalization.
\newblock \emph{arXiv preprint arXiv:1803.05407}, 2018.

\bibitem[Izmailov et~al.(2021)Izmailov, Vikram, Hoffman, and Wilson]{izmailov2021bayesian}
Pavel Izmailov, Sharad Vikram, Matthew~D Hoffman, and Andrew Gordon~Gordon Wilson.
\newblock What are bayesian neural network posteriors really like?
\newblock In \emph{International conference on machine learning}, pages 4629--4640. PMLR, 2021.

\bibitem[Jospin et~al.(2022)Jospin, Laga, Boussaid, Buntine, and Bennamoun]{jospin2022hands}
Laurent~Valentin Jospin, Hamid Laga, Farid Boussaid, Wray Buntine, and Mohammed Bennamoun.
\newblock Hands-on bayesian neural networks—a tutorial for deep learning users.
\newblock \emph{IEEE Computational Intelligence Magazine}, 17\penalty0 (2):\penalty0 29--48, 2022.

\bibitem[Jungo and Reyes(2019)]{jungo2019assessing}
Alain Jungo and Mauricio Reyes.
\newblock Assessing reliability and challenges of uncertainty estimations for medical image segmentation.
\newblock In \emph{International Conference on Medical Image Computing and Computer-Assisted Intervention}, pages 48--56. Springer, 2019.

\bibitem[Jungo et~al.(2020)Jungo, Balsiger, and Reyes]{jungo2020analyzing}
Alain Jungo, Fabian Balsiger, and Mauricio Reyes.
\newblock Analyzing the quality and challenges of uncertainty estimations for brain tumor segmentation.
\newblock \emph{Frontiers in neuroscience}, page 282, 2020.

\bibitem[Kapoor et~al.(2022)Kapoor, Maddox, Izmailov, and Wilson]{kapoor2022uncertainty}
Sanyam Kapoor, Wesley~J Maddox, Pavel Izmailov, and Andrew~Gordon Wilson.
\newblock On uncertainty, tempering, and data augmentation in bayesian classification.
\newblock \emph{arXiv preprint arXiv:2203.16481}, 2022.

\bibitem[Kendall and Gal(2017)]{kendall2017uncertainties}
Alex Kendall and Yarin Gal.
\newblock What uncertainties do we need in bayesian deep learning for computer vision?
\newblock \emph{Advances in neural information processing systems}, 30, 2017.

\bibitem[Kendall et~al.(2015)Kendall, Badrinarayanan, and Cipolla]{kendall2015bayesian}
Alex Kendall, Vijay Badrinarayanan, and Roberto Cipolla.
\newblock Bayesian segnet: Model uncertainty in deep convolutional encoder-decoder architectures for scene understanding.
\newblock \emph{arXiv preprint arXiv:1511.02680}, 2015.

\bibitem[Kohl et~al.(2018)Kohl, Romera-Paredes, Meyer, De~Fauw, Ledsam, Maier-Hein, Eslami, Rezende, and Ronneberger]{kohl2018probabilistic}
Simon~AA Kohl, Bernardino Romera-Paredes, Clemens Meyer, Jeffrey De~Fauw, Joseph~R Ledsam, Klaus~H Maier-Hein, SM~Eslami, Danilo~Jimenez Rezende, and Olaf Ronneberger.
\newblock A probabilistic u-net for segmentation of ambiguous images.
\newblock \emph{arXiv preprint arXiv:1806.05034}, 2018.

\bibitem[Lakshminarayanan et~al.(2016)Lakshminarayanan, Pritzel, and Blundell]{lakshminarayanan2016simple}
Balaji Lakshminarayanan, Alexander Pritzel, and Charles Blundell.
\newblock Simple and scalable predictive uncertainty estimation using deep ensembles.
\newblock \emph{arXiv preprint arXiv:1612.01474}, 2016.

\bibitem[Lampinen and Vehtari(2001)]{lampinen2001bayesian}
Jouko Lampinen and Aki Vehtari.
\newblock Bayesian approach for neural networks—review and case studies.
\newblock \emph{Neural networks}, 14\penalty0 (3):\penalty0 257--274, 2001.

\bibitem[Larrazabal et~al.(2021)Larrazabal, Mart{\'\i}nez, Dolz, and Ferrante]{larrazabal2021orthogonal}
Agostina~J Larrazabal, C{\'e}sar Mart{\'\i}nez, Jose Dolz, and Enzo Ferrante.
\newblock Orthogonal ensemble networks for biomedical image segmentation.
\newblock In \emph{International Conference on Medical Image Computing and Computer-Assisted Intervention}, pages 594--603. Springer, 2021.

\bibitem[Lifshitz and Landau(1984)]{lifshitz1984statistical}
EM~Lifshitz and Lev~Davidovich Landau.
\newblock Statistical physics (course of theoretical physics, volume 5), 1984.

\bibitem[MacKay(1995)]{mackay1995bayesian}
David~JC MacKay.
\newblock Bayesian neural networks and density networks.
\newblock \emph{Nuclear Instruments and Methods in Physics Research Section A: Accelerators, Spectrometers, Detectors and Associated Equipment}, 354\penalty0 (1):\penalty0 73--80, 1995.

\bibitem[Maddox et~al.(2019)Maddox, Izmailov, Garipov, Vetrov, and Wilson]{maddox2019simple}
Wesley~J Maddox, Pavel Izmailov, Timur Garipov, Dmitry~P Vetrov, and Andrew~Gordon Wilson.
\newblock A simple baseline for bayesian uncertainty in deep learning.
\newblock \emph{Advances in Neural Information Processing Systems}, 32, 2019.

\bibitem[Marcot and Penman(2019)]{marcot2019advances}
Bruce~G Marcot and Trent~D Penman.
\newblock Advances in bayesian network modelling: Integration of modelling technologies.
\newblock \emph{Environmental modelling \& software}, 111:\penalty0 386--393, 2019.

\bibitem[Mehrtash et~al.(2020)Mehrtash, Wells, Tempany, Abolmaesumi, and Kapur]{mehrtash2020confidence}
Alireza Mehrtash, William~M Wells, Clare~M Tempany, Purang Abolmaesumi, and Tina Kapur.
\newblock Confidence calibration and predictive uncertainty estimation for deep medical image segmentation.
\newblock \emph{IEEE transactions on medical imaging}, 39\penalty0 (12):\penalty0 3868--3878, 2020.

\bibitem[Messroghli et~al.(2004)Messroghli, Radjenovic, Kozerke, Higgins, Sivananthan, and Ridgway]{messroghli2004modified}
Daniel~R Messroghli, Aleksandra Radjenovic, Sebastian Kozerke, David~M Higgins, Mohan~U Sivananthan, and John~P Ridgway.
\newblock Modified look-locker inversion recovery (molli) for high-resolution t1 mapping of the heart.
\newblock \emph{Magnetic Resonance in Medicine: An Official Journal of the International Society for Magnetic Resonance in Medicine}, 52\penalty0 (1):\penalty0 141--146, 2004.

\bibitem[Minderer et~al.(2021)Minderer, Djolonga, Romijnders, Hubis, Zhai, Houlsby, Tran, and Lucic]{minderer2021revisiting}
Matthias Minderer, Josip Djolonga, Rob Romijnders, Frances Hubis, Xiaohua Zhai, Neil Houlsby, Dustin Tran, and Mario Lucic.
\newblock Revisiting the calibration of modern neural networks.
\newblock \emph{Advances in Neural Information Processing Systems}, 34:\penalty0 15682--15694, 2021.

\bibitem[Mingard et~al.(2021)Mingard, Valle-P{\'e}rez, Skalse, and Louis]{mingard2021sgd}
Chris Mingard, Guillermo Valle-P{\'e}rez, Joar Skalse, and Ard~A Louis.
\newblock Is sgd a bayesian sampler? well, almost.
\newblock \emph{Journal of Machine Learning Research}, 22, 2021.

\bibitem[Nabarro et~al.(2021)Nabarro, Ganev, Garriga-Alonso, Fortuin, van~der Wilk, and Aitchison]{nabarro2021data}
Seth Nabarro, Stoil Ganev, Adria Garriga-Alonso, Vincent Fortuin, Mark van~der Wilk, and Laurence Aitchison.
\newblock Data augmentation in bayesian neural networks and the cold posterior effect.
\newblock \emph{arXiv preprint arXiv:2106.05586}, 2021.

\bibitem[Neal(2012)]{neal2012bayesian}
Radford~M Neal.
\newblock \emph{Bayesian learning for neural networks}, volume 118.
\newblock Springer Science \& Business Media, 2012.

\bibitem[Neal et~al.(2011)]{neal2011mcmc}
Radford~M Neal et~al.
\newblock Mcmc using hamiltonian dynamics.
\newblock \emph{Handbook of markov chain monte carlo}, 2\penalty0 (11):\penalty0 2, 2011.

\bibitem[Ouyang et~al.(2022)Ouyang, Wang, Chen, Li, Bai, Kainz, and Rueckert]{ouyang2022improved}
Cheng Ouyang, Shuo Wang, Chen Chen, Zeju Li, Wenjia Bai, Bernhard Kainz, and Daniel Rueckert.
\newblock Improved post-hoc probability calibration for out-of-domain mri segmentation.
\newblock In \emph{International Workshop on Uncertainty for Safe Utilization of Machine Learning in Medical Imaging}, pages 59--69. Springer, 2022.

\bibitem[Ovadia et~al.(2019)Ovadia, Fertig, Ren, Nado, Sculley, Nowozin, Dillon, Lakshminarayanan, and Snoek]{ovadia2019can}
Yaniv Ovadia, Emily Fertig, Jie Ren, Zachary Nado, David Sculley, Sebastian Nowozin, Joshua~V Dillon, Balaji Lakshminarayanan, and Jasper Snoek.
\newblock Can you trust your model's uncertainty? evaluating predictive uncertainty under dataset shift.
\newblock \emph{arXiv preprint arXiv:1906.02530}, 2019.

\bibitem[Risten(1989)]{risten1989fokker}
H~Risten.
\newblock The fokker-planck equation: methods of solution and applications.
\newblock \emph{Springer Series in Synergetics}, 18:\penalty0 544--3, 1989.

\bibitem[Ronneberger et~al.(2015)Ronneberger, Fischer, and Brox]{ronneberger2015u}
Olaf Ronneberger, Philipp Fischer, and Thomas Brox.
\newblock U-net: Convolutional networks for biomedical image segmentation.
\newblock In \emph{International Conference on Medical image computing and computer-assisted intervention}, pages 234--241. Springer, 2015.

\bibitem[Salahuddin et~al.(2023)Salahuddin, Chen, Zhong, Woodruff, Rad, Mali, and Lambin]{salahuddin2023head}
Zohaib Salahuddin, Yi~Chen, Xian Zhong, Henry~C Woodruff, Nastaran~Mohammadian Rad, Shruti~Atul Mali, and Philippe Lambin.
\newblock From head and neck tumour and lymph node segmentation to survival prediction on pet/ct: An end-to-end framework featuring uncertainty, fairness, and multi-region multi-modal radiomics.
\newblock \emph{Cancers}, 15\penalty0 (7):\penalty0 1932, 2023.

\bibitem[S{\"a}rkk{\"a} and Solin(2019)]{sarkka2019applied}
Simo S{\"a}rkk{\"a} and Arno Solin.
\newblock \emph{Applied stochastic differential equations}, volume~10.
\newblock Cambridge University Press, 2019.

\bibitem[Speagle(2019)]{speagle2019conceptual}
Joshua~S Speagle.
\newblock A conceptual introduction to markov chain monte carlo methods.
\newblock \emph{arXiv preprint arXiv:1909.12313}, 2019.

\bibitem[Wang et~al.(2021)Wang, Feng, and Zhang]{wang2021rethinking}
Deng-Bao Wang, Lei Feng, and Min-Ling Zhang.
\newblock Rethinking calibration of deep neural networks: Do not be afraid of overconfidence.
\newblock \emph{Advances in Neural Information Processing Systems}, 34:\penalty0 11809--11820, 2021.

\bibitem[Wang et~al.(2019)Wang, Li, Aertsen, Deprest, Ourselin, and Vercauteren]{wang2019aleatoric}
Guotai Wang, Wenqi Li, Michael Aertsen, Jan Deprest, S{\'e}bastien Ourselin, and Tom Vercauteren.
\newblock Aleatoric uncertainty estimation with test-time augmentation for medical image segmentation with convolutional neural networks.
\newblock \emph{Neurocomputing}, 338:\penalty0 34--45, 2019.

\bibitem[Wang and Yeung(2020)]{wang2020survey}
Hao Wang and Dit-Yan Yeung.
\newblock A survey on bayesian deep learning.
\newblock \emph{ACM Computing Surveys (CSUR)}, 53\penalty0 (5):\penalty0 1--37, 2020.

\bibitem[Wenzel et~al.(2020)Wenzel, Roth, Veeling, Swiatkowski, Tran, Mandt, Snoek, Salimans, Jenatton, and Nowozin]{wenzel2020good}
Florian Wenzel, Kevin Roth, Bastiaan~S Veeling, Jakub Swiatkowski, Linh Tran, Stephan Mandt, Jasper Snoek, Tim Salimans, Rodolphe Jenatton, and Sebastian Nowozin.
\newblock How good is the bayes posterior in deep neural networks really?
\newblock \emph{arXiv preprint arXiv:2002.02405}, 2020.

\bibitem[Yan et~al.(2020)Yan, Huang, Xia, Gu, and Tao]{2020MRI}
W.~Yan, L.~Huang, L.~Xia, S.~Gu, and Q.~Tao.
\newblock Mri manufacturer shift and adaptation: Increasing the generalizability of deep learning segmentation for mr images acquired with different scanners.
\newblock 2020.

\bibitem[Yan et~al.(2019)Yan, Wang, Gu, Huang, Yan, Xia, and Tao]{yan2019domain}
Wenjun Yan, Yuanyuan Wang, Shengjia Gu, Lu~Huang, Fuhua Yan, Liming Xia, and Qian Tao.
\newblock The domain shift problem of medical image segmentation and vendor-adaptation by unet-gan.
\newblock In \emph{Medical Image Computing and Computer Assisted Intervention--MICCAI 2019: 22nd International Conference, Shenzhen, China, October 13--17, 2019, Proceedings, Part II 22}, pages 623--631. Springer, 2019.

\bibitem[Zhang et~al.(2019)Zhang, Li, Zhang, Chen, and Wilson]{zhang2019cyclical}
Ruqi Zhang, Chunyuan Li, Jianyi Zhang, Changyou Chen, and Andrew~Gordon Wilson.
\newblock Cyclical stochastic gradient mcmc for bayesian deep learning.
\newblock \emph{arXiv preprint arXiv:1902.03932}, 2019.

\bibitem[Zhao et~al.(2022)Zhao, Yang, Schweidtmann, and Tao]{zhao2022efficient}
Yidong Zhao, Changchun Yang, Artur Schweidtmann, and Qian Tao.
\newblock Efficient bayesian uncertainty estimation for nnu-net.
\newblock In \emph{International Conference on Medical Image Computing and Computer-Assisted Intervention}, pages 535--544. Springer, 2022.

\end{thebibliography}
